\documentclass[twocolumn,twocolappendix]{aastex631}

\usepackage{listings}
\usepackage{tabularx}
\usepackage{graphicx}
\usepackage{epstopdf}
\usepackage{enumitem}

\shorttitle{MaNGA Merger Stages with ML}
\shortauthors{Chang et al.}

\begin{document}

\title{SDSS-IV MaNGA: Unveiling Galaxy Interaction by Merger Stages with Machine Learning}

\correspondingauthor{Yu-Yen~Chang}
\email{yuyenchang.astro@gmail.com}

\author[0000-0002-6720-8047]{Yu-Yen~Chang}
\affiliation{Department of Physics, National Chung Hsing University, 40227, Taichung, Taiwan}
\affiliation{Academia Sinica Institute of Astronomy and Astrophysics, Taipei 10617, Taiwan}
\author[0000-0001-7218-7407]{Lihwai~Lin}
\affiliation{Academia Sinica Institute of Astronomy and Astrophysics, Taipei 10617, Taiwan}
\author[0000-0002-1370-6964]{Hsi-An~Pan}
\affiliation{Department of Physics, Tamkang University, No.151, Yingzhuan Road, Tamsui District, New Taipei City 251301, Taiwan}
\author[0000-0003-3512-6204]{Chieh-An~Lin}
\affiliation{WPC Systems Ltd., 55 Qingyun Road, Tucheng, New Taipei City, 23653, Taiwan}
\author[0000-0001-5615-4904]{Bau-Ching~Hsieh}
\affiliation{Academia Sinica Institute of Astronomy and Astrophysics, Taipei 10617, Taiwan}
\author[0000-0003-4758-4501]{Connor Bottrell}
\affiliation{Kavli Institute for the Physics and Mathematics of the Universe (WPI), UTIAS, University of Tokyo, Kashiwa, Chiba 277-8583, Japan}
\author{Pin-Wei~Wang}
\affiliation{National Museum of Natural Science, No. 1, Guanqian Rd, North District, Taichung, 404, Taiwan}



\begin{abstract}

We use machine learning techniques to classify galaxy merger stages, which can unveil physical processes that drive the star formation and active galactic nucleus (AGN) activities during galaxy interaction. The sample contains 4,690 galaxies from the integral field spectroscopy survey SDSS-IV MaNGA, and can be separated to 1,060 merging galaxies and 3630 non-merging or unclassified galaxies. For the merger sample, there are 468, 125, 293, and 174 galaxies in (1) incoming pair phase, (2) first pericentric passage phase, (3) aproaching or just passing the apocenter, and (4) final coalescence phase or post-mergers. With the information of projected separation, line-of-sight velocity difference, SDSS $gri$ images, and MaNGA H$\alpha$ velocity map, we are able to classify the mergers and their stages with good precision, which is the most important score to identify interacting galaxies. For the 2-phase classification (binary; non-merger and merger), the performance can be high (precision$>$0.90) with \texttt{LGBMClassifier}. We find that sample size can be increased by rotation, so the 5-phase classification (non-merger, 1, 2, 3, and 4 merger stages) can be also good (precision$>$0.85).  The most important features come from SDSS $gri$ images.  The contribution from MaNGA H$\alpha$ velocity map,  projected separation, and line-of-sight velocity difference can further improve the performance by 0-20\%.  In other words, the image and the velocity information are sufficient to capture important features of galaxy interactions,  and our results can apply to the entire MaNGA data as well as future all-sky surveys. 

\end{abstract}



\section{Introduction} \label{sec1}

It is well known that galaxy interaction is one of the key drivers of galaxy evolution. In the structural models, galaxies assembled their masses through merging and accretion at the center of dark matter halos \citep[e.g., ][]{1978MNRAS.183..341W,1991ApJ...379...52W,2015ARA&A..53...51S}. Observationally, galaxies are found to evolve over cosmic time from star-forming to quiescent phase \citep[e.g.,][]{2007ApJ...665..265F,2010ApJ...721..193P,2012ApJ...753..167B}, and galaxy interactions play an important role in galaxy assembly, morphological transformation, and central black hole growth \citep[e.g.,][]{2000MNRAS.311..576K,2003ApJ...597..893N,2006ApJS..163....1H,2014ARA&A..52..291C}. 

Earlier statistical studies investigating the impact of mergers on galaxy properties have been primarily based on single-fiber  or single-slit spectroscopic samples, which rely on either only the information in the central parts of galaxies or integrated properties \citep[e.g., ][]{2004MNRAS.355..874N, 2007ApJ...660L..51L,2008AJ....135.1877E,2010AJ....139.1857W,2012MNRAS.426..549S,2013MNRAS.433L..59P,2015MNRAS.454.1742K}. On the other hand, the Integral field spectroscopy (IFS) surveys, such as ATLAS-3D \citep{2011MNRAS.413..813C}, Calar Alto Legacy Integral Field Area \citep[CALIFA,][]{2012A&A...538A...8S}, Sydney-AAO Multi-object IFS survey \citep[SAMI,][]{2015MNRAS.447.2857B}, Mapping Nearby Galaxies at Apache Point Observatory \citep[MaNGA,][]{2015ApJ...798....7B}, and HECTOR \citep{2016SPIE.9908E..1FB}, offer statistical numbers of galaxies with spatially resolved measurements, allowing for the studies of the spatial distributions of star formation activities and kinematics \citep[e.g.,][]{2014A&A...562A..47G,2017ApJ...851L..24H,2018MNRAS.474.2039E,2019ApJ...884L..33L}. These surveys, therefore, provide promising opportunities to investigate the effect of mergers in greater detail. 

Star formation activity potentially varies from stage to stage \citep[e.g, ][]{2005MNRAS.361..776S, 2008ApJS..175..356H, 2015MNRAS.446..521S, 2015MNRAS.449...49R}, and therefore it is desirable to classify merger stage in observation to test the simulation predictions.  The effect of mergers and merger stages has been investigated with both spectroscopically identified pairs and visually identified post-mergers \citep[e.g., ][]{2008AJ....135.1877E, 2013MNRAS.435.3627E, 2004MNRAS.355..874N, 2012MNRAS.426..549S, 2014MNRAS.437.2137S, 2015MNRAS.454.1742K, 2019MNRAS.482L..55T}.  The observational evidence is consistent with the theoretical picture of gas inflows and centrally-enhanced star-formation and gas metallicity depletion. Several non-parametric morphological parameters have been suggested as a means of identifying mergers \citep[e.g.,][]{1994ApJ...432...75A, 2003ApJS..147....1C, 2004AJ....128..163L, 2019MNRAS.483.4140R}, but they could suffer from some degree of incompleteness and impurity stemming from the fact that the information in the images or maps is being condensed to few parameters and that much information is lost in this process. \citet{2019ApJ...872...76N} used \texttt{GADGET-3/SUNRISE} hydrodynamic simulations of merging galaxies and linear discriminant analysis to produce merging galaxies, and adapted the simulated images to the specifications of the Sloan Digital Sky Survey \citep[SDSS,][]{2006AJ....131.2332G} imaging. They created an accurate merging galaxy classifier from imaging predictors,  which has potential to reveal more complete merger samples from imaging and IFS surveys \citep{2021ApJ...912...45N}.  
\citet{2019ApJ...881..119P} made a first attempt to make the comparison between star formation and pair separation using visually classified merger stages. They classified galaxy merger stages via visual inspections using the composite images observed by the 2.5 m Telescope of the SDSS to four merger stages.  They also presented an empirical picture of spatially resolved interaction-triggered star formation rate (SFR) as a function of merger sequence using the IFS data from the MaNGA survey.  However, visual classification is quite time-consuming \citep[e.g.,][]{2008MNRAS.389.1179L}, but inspection over a large sample is necessary because galaxy interactions have diverse appearance and configurations. 

Recently,  machine learning (ML) has been applied to derive various physical parameters of galaxies \citep[e.g.,][]{2015ApJ...813...53M,2016A&A...596A..39K,2018A&A...609A.111D,2019ApJ...881L..14H,2019MNRAS.489.4817D,2019A&A...622A.137B, 2021ApJ...920...68C}, and improves upon linear combinations through non-linear activations \citep[e.g.,][]{2018MNRAS.479..415A, 2019MNRAS.483.2968W, 2021MNRAS.504..372B, 2022MNRAS.514.3294B, 2020ApJ...895..115F, 2022ApJ...931...34F}. In particular, classification by ML \citep[e.g., ][]{2010MNRAS.406..342B,2015ApJS..221....8H,2018MNRAS.476.3661D,2019MNRAS.490.5390B, 2019A&A...626A..49P,2020A&C....3000334B,2021ApJ...920...68C} can avoid time-consuming visual inspections, and will be helpful for the visual classification of galaxy-galaxy interactions from the forthcoming large surveys. For instance, \citet{2019A&A...626A..49P} developed a convolutional neural network (CNN) architecture, which was with observational SDSS and simulated EAGLE images to identify galaxy mergers. They showed that the networks achieve better performance in observational data than in simulations. \citet{2020ApJ...895..115F} achieve 0.90 accuracy to classify major mergers and measure galaxy merger in all five CANDELS fields using CNN trained with simulated galaxies from the IllustrisTNG simulation,  and separate star-forming galaxies from post-mergers in a following work. \citet{2022ApJ...931...34F}.  \citet{2022MNRAS.514.3294B} deployed a CNN and evaluated on mock observations of simulated galaxies from the IllustrisTNG simulations to identify post-mergers.  \citet{2022MNRAS.511..100B} examine both the morphological and kinematic features of merger remnants from the TNG100, and show that the stellar kinematic data have little contributions. Moreover, it has been discussed whether ancillary information such as kinematics and spectroscopic information to the images may provide an additional basis for classification \citep[e.g., ][]{2019ApJ...881..119P, 2019ApJ...872...76N, 2022MNRAS.tmp.1734M, 2022MNRAS.511..100B}. Therefore, it is important to identify specific features for classification with ML both from photometric and spectroscopic observables.

In this paper, we will compare several algorithms from \texttt{scikit-learn} and \texttt{XGBoost}, and show our results by using the state-of-the-art ML methods to identify merger stages for the MaNGA sample.
The structure of this paper is as follows.
We describe the data and sample selections in Section \ref{sec2}.
We analyze the properties in Section \ref{sec3}.
We discuss the results in Section \ref{sec4} and summarize in Section \ref{sec5}.
Throughout the paper, we use AB magnitudes, adopt the cosmological parameters
($\Omega_{\rm M}$,$\Omega_\Lambda$,$h$)=(0.30,0.70,0.70), and assume the
stellar initial mass function of \citet{2003PASP..115..763C}.
\section{Data} \label{sec2}

\begin{figure}
\centering
\includegraphics[width=1.00\columnwidth]{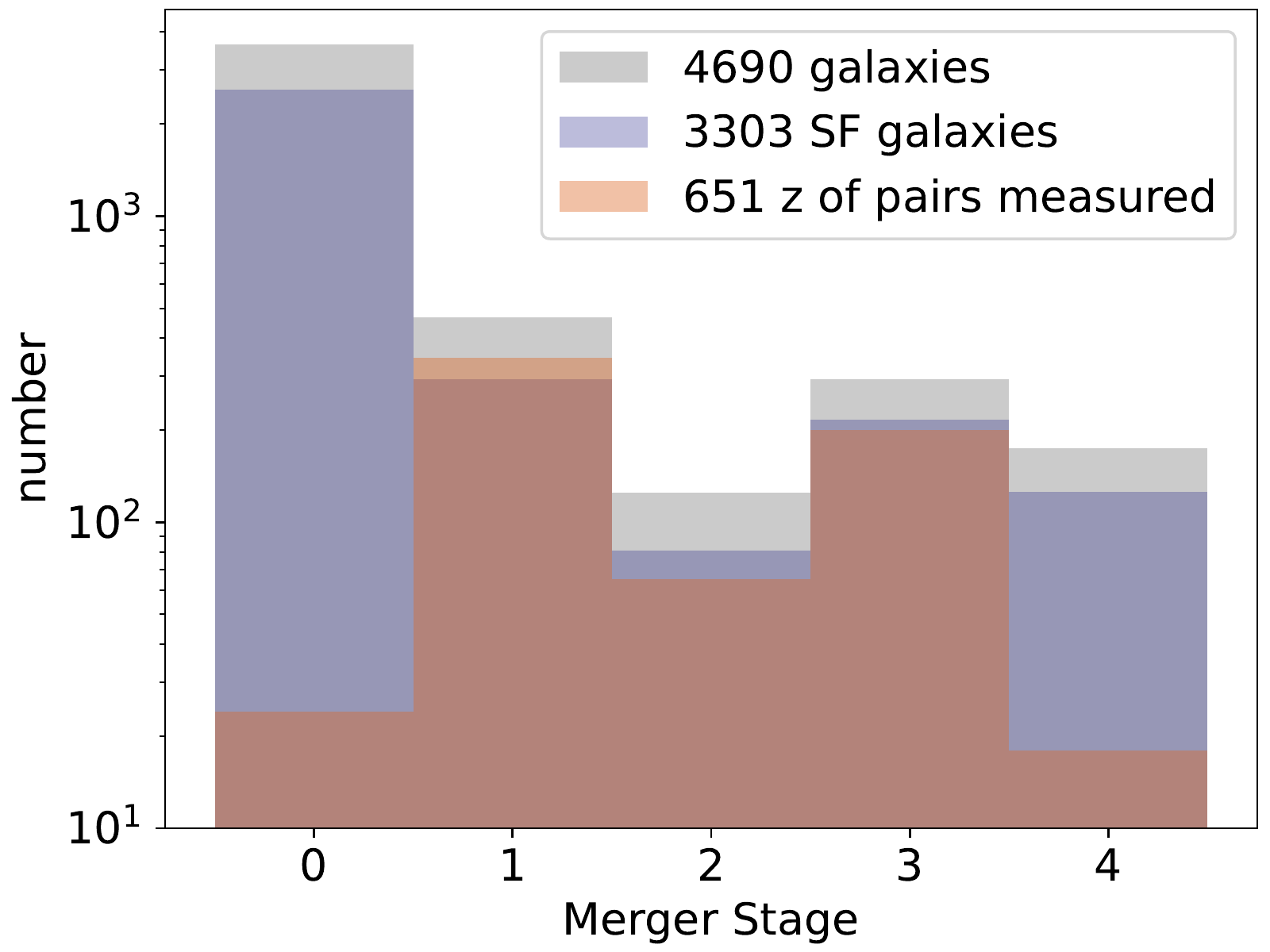} 
\caption{The histogram of the merger stages. The gray color shows the whole sample (N=4,690), the blue color shows the star-forming galaxies (N=3,303), and the red color shows the objects which contains redshift of two members measured and provide} information of projected separation and the difference in the line-of-sight velocity (N=651).
\label{stages}
\end{figure}

\begin{figure*}
\centering
\includegraphics[width=0.90\textwidth]{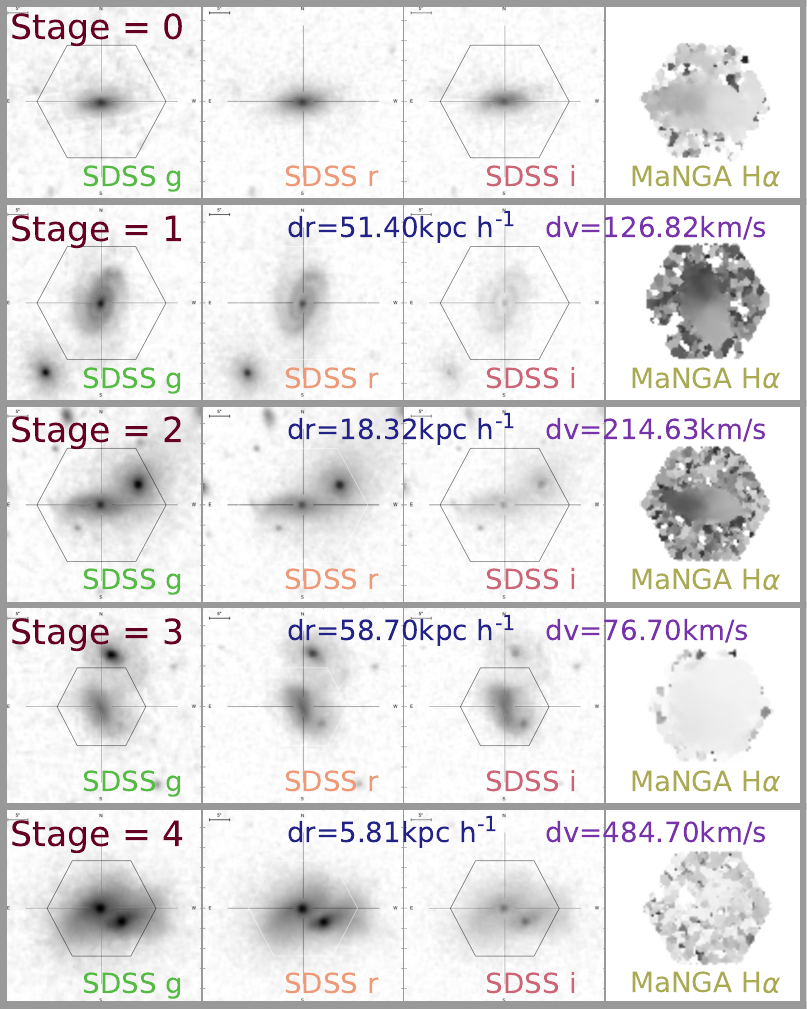} 
\caption{Example of merger stages with input data for classification: the projected separation (dr in kpc $\rm{h}^{-1}$),  the velocity difference (dv in  km/s), the SDSS $gri$-band image, and the MaNGA H$\alpha$ velocity map.}
\label{images}
\end{figure*}

MaNGA \citep{2015ApJ...798....7B,2016AJ....151....8Y,2016AJ....152..197Y,2017AJ....154...86W} is an integral field unit (IFU) survey on the SDSS 2.5 m telescope \citep{2006AJ....131.2332G}, as part of the SDSS-IV survey \citep{2017ApJS..233...25A,2017AJ....154...28B}. MaNGA makes use of a modification of the BOSS spectrographs \citep{2013AJ....146...32S} to bundle fibers into hexagons \citep{2015AJ....149...77D}.  Each spectrum has a wavelength coverage of 3,500-10,000 and instrumental resolution about 60$km/s$. After dithering, MaNGA data have an effective spatial resolution of 2$\arcsec$.5 \citep[][]{2015AJ....150...19L}, and datacubes are gridded with 0$\arcsec$.5 spaxels. In this work, we use 4,690 galaxies with major-to-minor-axis ratio greater than 0.4 at z$<$0.15 from the MaNGA DR15 version as described in \citet{2019ApJ...881..119P}.

Among these 4,690 sample, there are 1,060 merger galaxies and 3,630 non-interacting galaxies or unclassified galaxies according to the classification in \citet{2019ApJ...881..119P}. Interactions between galaxies are classified according to the following merger stages:
\begin{itemize}
\item Stage 0: non-interacting galaxies,  unclassified galaxies, or a potential paired galaxy but without spec-z confirmation for its companion.
\item Stage 1: well-separated pairs which do not show any morphology distortion, i.e., incoming pairs, before the first pericenter passage.
\item Stage 2: close pairs showing strong signs of interaction, such as tails and bridges, i.e., after the first pericenter passage.
\item Stage 3: well-separated pairs, but showing weak morphology, distortion, i.e., approaching the apocenter or just, passing the apocenter.
\item Stage 4: two components strongly overlapping with each other and showing strong morphological distortion, i.e., final coalescence phase, or single galaxies with obvious tidal features such as tails and shells, i.e., post-mergers.
\end{itemize}

According to the visual inspection in \citet{2019ApJ...881..119P}, there are 468 galaxies in merger stage 1, 125 galaxies in merger stage 2, 293 galaxies in merger stage 3, and 174 galaxies in merger stage 4.  Therefore, there are 1,060 out of the 4,690 sample are classified as interaction galaxies in one of the stages as shown in Figure~\ref{stages}.  Among those, 651 objects contain NASA-Sloan Atlas (NSA) redshifts of two members (pairs) measured, which provide information of projected separation ($dr$) and the difference in the line-of-sight velocity ($dv$).  Out of these 651 systems, there are 344 galaxies in merger stage 1, 65 galaxies in merger stage 2, 200 galaxies in merger stage 3, and 18 galaxies in merger stage 4. We adopt the values of $dr$ in kpc $\rm{h}^{-1}$ and $dv$ in km/s as input data. For objects without $dr$ and $dv$, we use -999 as input values. As quiescent galaxies do not have prominent features during galaxy-galaxy interaction as star-forming galaxies, in this work, we only focus on the latter. Among the 4690 DR15 galaxuies, 3,303 objects can be selected as star-forming galaxies with $\log$(sSFR/$\rm{yr}^{-1}$)$>$-11 (e.g., \citealt{2011MNRAS.413..996M, 2013MNRAS.432..336W, 2012ApJ...754L..29W, 2014ApJ...795..104W, 2014ApJ...782...33L, 2015ApJ...801...80L, 2016ApJ...817..118T, 2018PASJ...70S..23J}; note that this can only be used in narrow redshift range as MaNGA and SDSS, \citealp[e.g., ][]{2019MNRAS.485.4817D}). In total, there are 294 galaxies in merger stage 1,  81 galaxies in merger stage 2, 216 galaxies in merger stage 3, and 126 galaxies in merger stage 4 satisfying this criterion. We include all 4,690 MaNGA galaxies in this paper, and limit our sample to 651 objects with redshifts of pairs measured and 3,303 star-forming galaxies for testing purpose.

The visual inspection by \citet{2019ApJ...881..119P} is mainly based on the projected separation ($dr$),  the velocity difference ($dv$), the SDSS $gri$-band image, and the MaNGA H$\alpha$ velocity map (for the purpose of redshift identification for the companion, instead of using the kinematic feature). In this paper, we also adopt the above information as input data as shown in Figure~\ref{images}. 

For SDSS images, 50$\arcsec\times$50$\arcsec$ combined images were adopted for visual classification in \citet{2019ApJ...881..119P}. Here we decompose these images to gri bands, each with 281 pixels $\times$ 281 pixels, resulting in 78,961 separate inputs. Meanwhile, for MaNGA H$\alpha$ velocity maps, the original image sizes varies from one to another. To be able to create unified datasets for our analysis, we resample each of them on a grid of 50 pixels $\times$ 50 pixels, such that an image contains 2,500 separate inputs. For the whole sample, there are 22 objects without any detection in the MaNGA H$\alpha$ velocity maps, so we choose -9999 for their spaxel values. For low signal-to-noise ratio objects, we still use the measurements from the MaNGA H$\alpha$ velocity maps to include all information. 


\section{Analysis} \label{sec3}

\subsection{Evaluations}

We evaluate the quality of the classification schemes with their performance.  First, we defined true positive (TP; a merger source which is classified as merger), true negative (TN: a non-merger source which is not classified as merger), false positive (FP: a non-merger source which is classified as merger), and false negative (FN: a merger source which is not classified as merger). Therefore, the true positive rate (TPR), the true negative rate (TNR), the false positive rate (FPR), and the false negative rate (FNR) are $TPR = TP/(TP+FN)$, $TNR = TN/(TN+FP)$, $FPR = FP/(TN+FP)$, and $FNR = FN/(TP+FN)$, respectively.  A good classification will categorize a merger as a merger (TPR) and a non-merger as a non-merger (TNR), rather than a non-merger as a merger (FPR) and a merger as a non-merger (FNR).  Accordingly, the result would reach high TPR, high TNR, low FPR, and low FNR.  Therefore, high performance can be determined by high accuracy, high precision, high recall, and high F1 score as described below. 

\begin{itemize}
\item Accuracy: fraction of sources (merger and non-merger) which are classified correctly over all sources. 
\begin{equation}
ACC=\frac{TP+TN}{TP+TN+FP+FN}
\end{equation}
\item Precision: merger sources which are classified correctly as mergers over all classified mergers. 
\begin{equation}
P=\frac{TP}{TP+FP}
\end{equation}
\item Recall: merger sources which are classified correctly as mergers over all merger sources. 
\begin{equation}
R=\frac{TP}{TP+FN}
\end{equation}
\item F1 score:  a harmonic mean of the precision and the recall. 
\begin{equation}
F1=2\times\frac{P\times R}{P+R}
\end{equation}
\end{itemize}

\subsection{Techniques and Parameters}

We use algorithms from Python packages, \texttt{scikit-learn} \footnote{http://scikit-learn.org/} and \texttt{XGBoost}\footnote{https://xgboost.readthedocs.io/}, to classify galaxy merger stages. In this subsection, we describe several classifiers which are commonly used and applied in this paper including \texttt{LGBMClassifier}, \texttt{LogisticRegression}, \texttt{DecisionTreeClassifier}, \texttt{RandomForestClassifier}, \texttt{KNeighborsClassifier}, \texttt{MLPClassifier}, \texttt{AdaBoostClassifier}, \texttt{GaussianNB}, and \texttt{XGBoost} \citep{scikit-learn,Chen:2016:XST:2939672.2939785}.

\begin{enumerate}

\item \texttt{LGBMClassifier}: a fast and high performance gradient boosting classifier which uses decision tree algorithms. It is a popular learning algorithm which can handle large sample size and takes lower memory. The input parameters are as in below.  
\begin{lstlisting}[language=Python]
boosting_type='gbdt', 
num_leaves=31, 
max_depth=- 1, 
learning_rate=0.1, 
n_estimators=100, 
subsample_for_bin=200000, 
objective=None, 
class_weight=None, 
min_split_gain=0.0, 
min_child_weight=0.001, 
min_child_samples=20, 
subsample=1.0, 
subsample_freq=0, 
colsample_bytree=1.0, 
reg_alpha=0.0, 
reg_lambda=0.0, 
random_state=None, 
n_jobs=- 1, 
importance_type='split'
\end{lstlisting}
We adopt mostly default parameters of the \texttt{LGBMClassifier} model. We test different combinations of input parameters, but did not find significant differences.  The dominant factors of the performance are the selection of input data and the choice of classifications as discussed in the following subsections. 

For other classifiers, we only describe input parameters which are different from default inputs. 

\item \texttt{LogisticRegression}: a common model for classification to estimate class probabilities by using a function to train a set of parameter.  Here we choose 1000 as the maximum number of iterations, `sag' solver due to the speed,  and 0.01 as the tolerance for stopping criteria.  Most cases converge before 100 epochs.

\item \texttt{DecisionTreeClassifier}: a predictive modeling classifier which uses decision trees as a predictive model to go from observations about an item represented in the branches to  conclude about the item's target value represented in the leaves. Here we choose 5 as the maximum depth of the tree.

\item \texttt{RandomForestClassifier}: an ensemble learning method for classification which use various decision tree classifiers. Here we choose 5 as the maximum depth, 10 as the number of trees in the forest, and 1 as the number of feature at each split.

\item \texttt{KNeighborsClassifier}: a non-parametric supervised learning method for classification with neighbor searches.  The input is the closest K training set, and the output is voted by its K nearest neighbors. Here we choose 3 as the K-neighbors of a point.

\item \texttt{MLPClassifier}:  a multi-layer perceptron (MLP) classifier by using neural network. Here we choose 1 as alpha (the strength of the L2 regularization term), and 1000 as the maximum number of iterations. There is one input layer, one hidden layer with 100 neurons, and one output layer. The activation function for the hidden layer is ’relu’, which is the rectified linear unit function. The training stops when the training loss does not improve by more than tol=0.0001 for n\_iter\_n\_change=10  consecutive passes over the training set.  Most cases converge before 50 epochs.

\item \texttt{GaussianNB}: Gaussian naive Bayes, which is based on applying Bayes' theorem with an assumption of the continuous values associated with each class are distributed according to a Gaussian distribution.

\item \texttt{AdaBoostClassifier}: adaptive boosting classifiers, which can be used in conjunction with many other types of learning algorithms.  Here the base estimator is \texttt{DecisionTreeClassifier} initialized with maximum depth equal to 1. We choose 50 as the maximum number of estimators at which boosting is terminated.

\item \texttt{XGBoost}: extreme gradient boosting, which is a machine learning algorithms to optimize distributed gradient boosting library and provides a parallel tree boosting learning. Here we use 5 as early stopping rounds which stop while lack of improvement,  and 100 number of round of iteration.  We choose 10 as the maximum depth of a tree, and 0.3 as `eta', which is the step size shrinkage used in update to prevent overfitting. We set the `objective' to `multi:softprob' to do multiclass classification using the softprob objective, which contains predicted probability of each data point belonging to each class. Most cases converge before 50 epochs.

\end{enumerate}

\subsection{Performances}\label{sec33}

We compare accuracy, precision, recall, and F1 score for different classifiers in Table~\ref{tab1}. Three classification cases are discussed as following.
\begin{itemize}
\item 5-phase classification ($P_5$)
	\begin{itemize}
	\item $P_5^0$: non-mergers 
	\item $P_5^1$: merger stage 1
	\item $P_5^2$: merger stage 2
	\item $P_5^3$: merger stage 3
	\item $P_5^4$: merger stage 4
	\end{itemize}
\item 3-phase classification($P_3$)
	\begin{itemize}
	\item $P_3^0$: non-mergers
	\item $P_3^1$: well-separated pairs:  merger stages 1 and 3
	\item $P_3^2$: very close pairs: merger stages 2 and 4
	\end{itemize}
\item 2-phase classification ($P_2$)
	\begin{itemize}
	\item $P_2^0$: non-mergers
	\item $P_2^1$: mergers: merger stages 1, 2, 3, and 4
	\end{itemize}
\end{itemize}

\begin{figure}
\centering
\includegraphics[width=1.00\columnwidth]{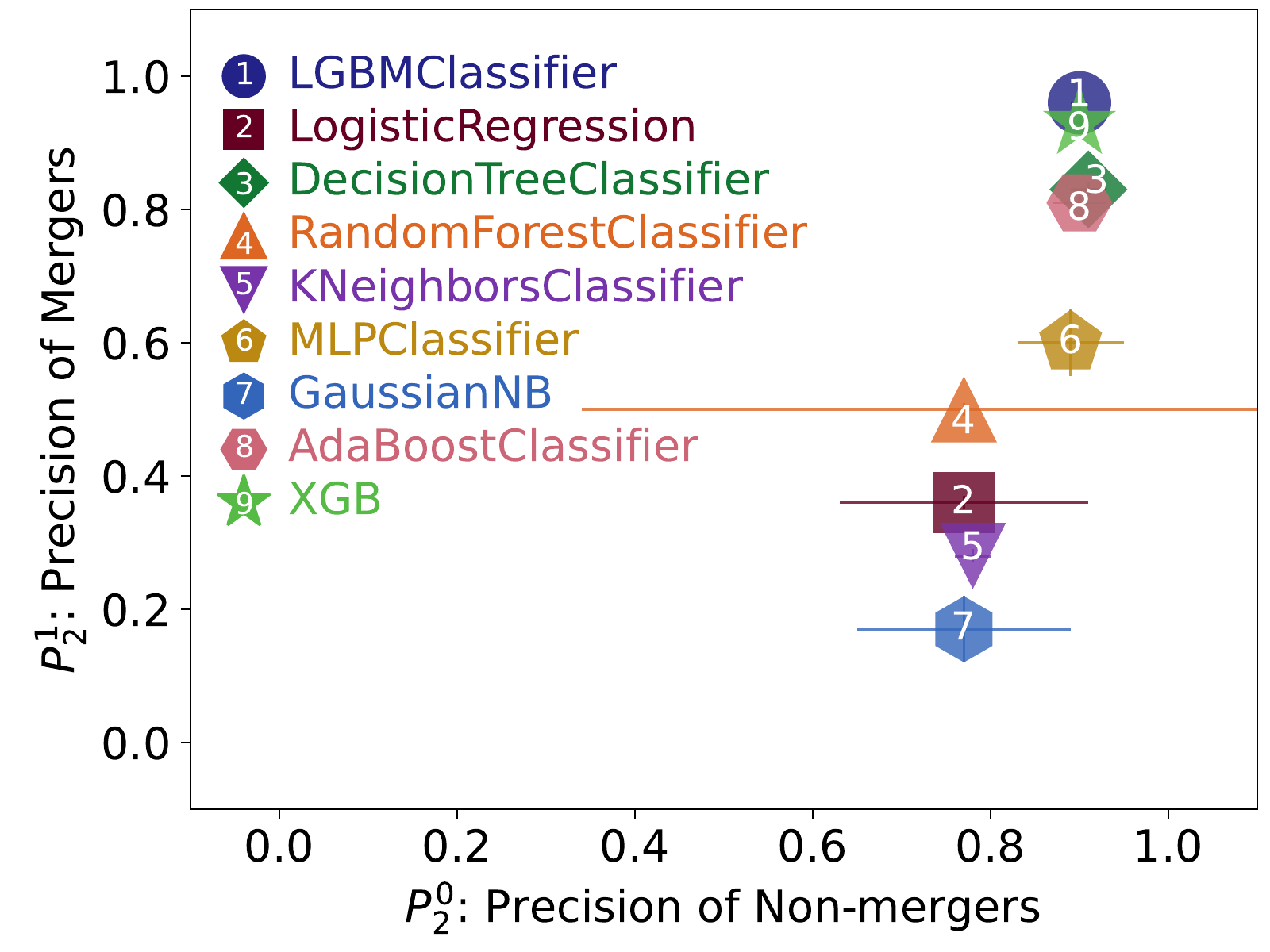} 
\caption{The precision of non-mergers and mergers of 2-phase classification for different classifiers with the original sample (N=4,690). }
\label{classifiers2}
\end{figure}

\begin{figure}
\centering
\includegraphics[width=1.00\columnwidth]{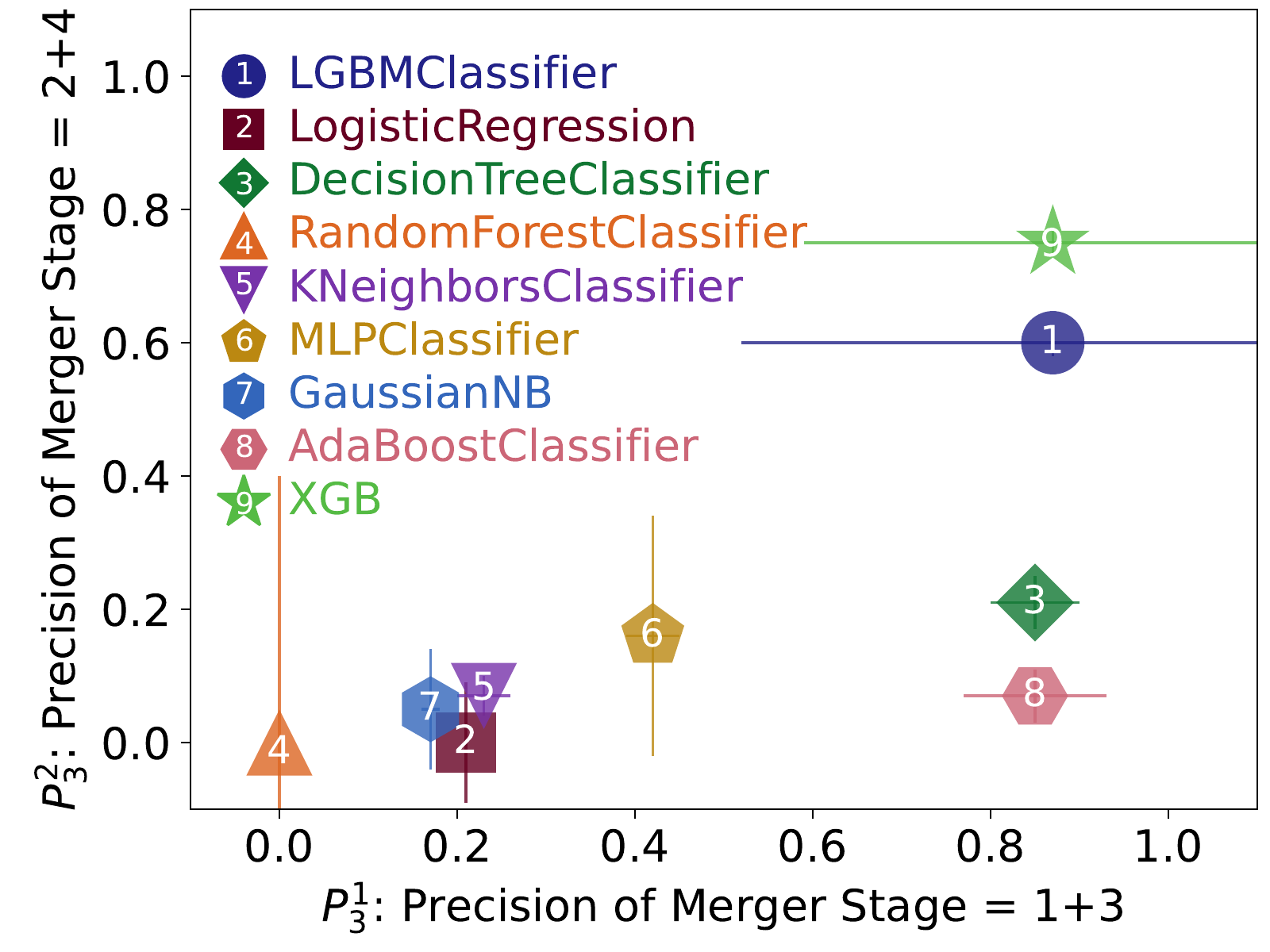} 
\caption{The precision of merger stage 1+3 and 2+4 of 3-phase classification for different classifiers with the original sample (N=4,690). }
\label{classifiers3}
\end{figure}

\begin{figure}
\centering
\includegraphics[width=1.00\columnwidth]{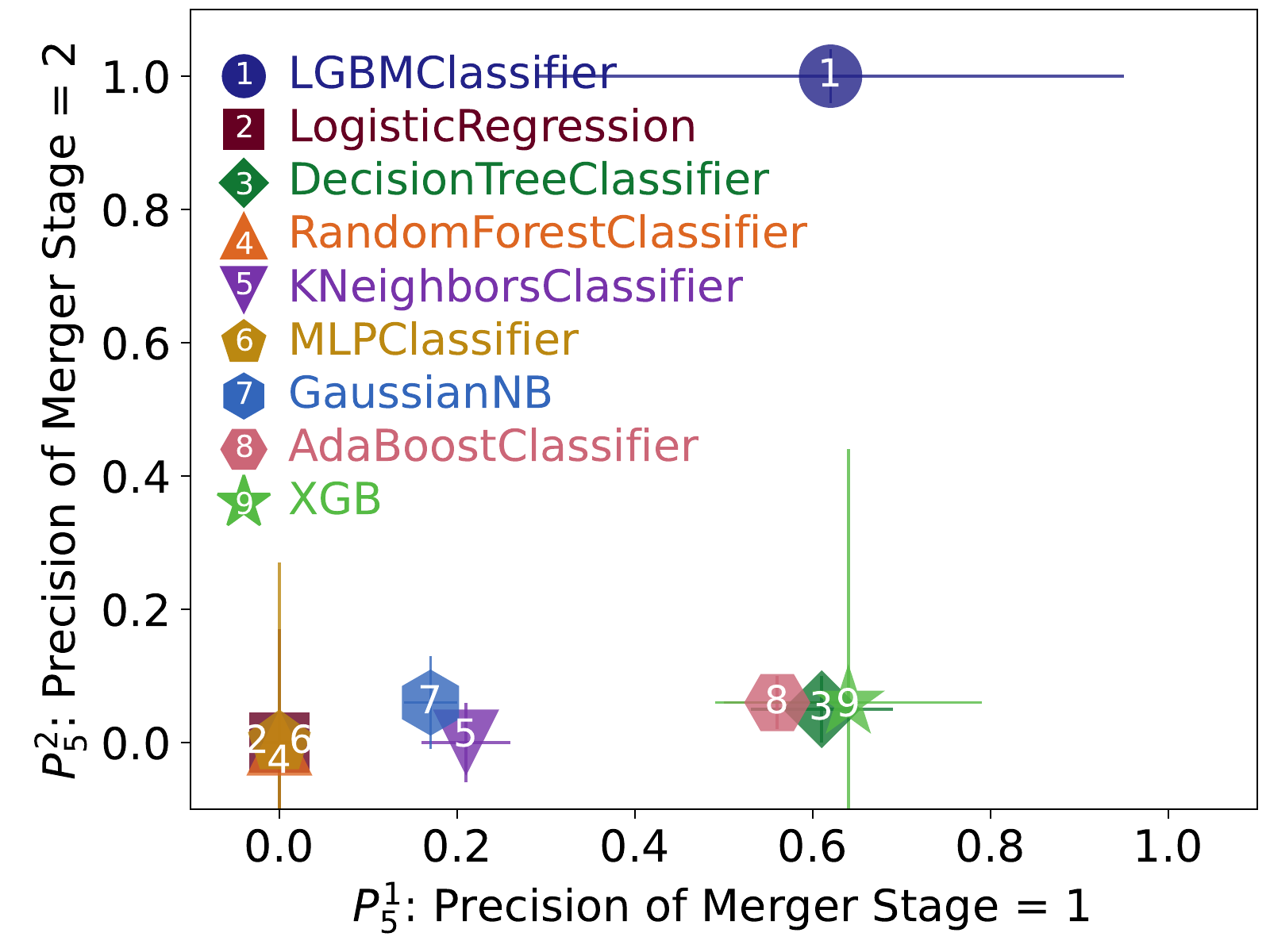} 
\caption{The precision of merger stage 1 and 2 of 5-phase classification for different classifiers with the original sample (N=4,690). }
\label{classifiers512}
\end{figure}

\begin{figure}
\centering
\includegraphics[width=1.00\columnwidth]{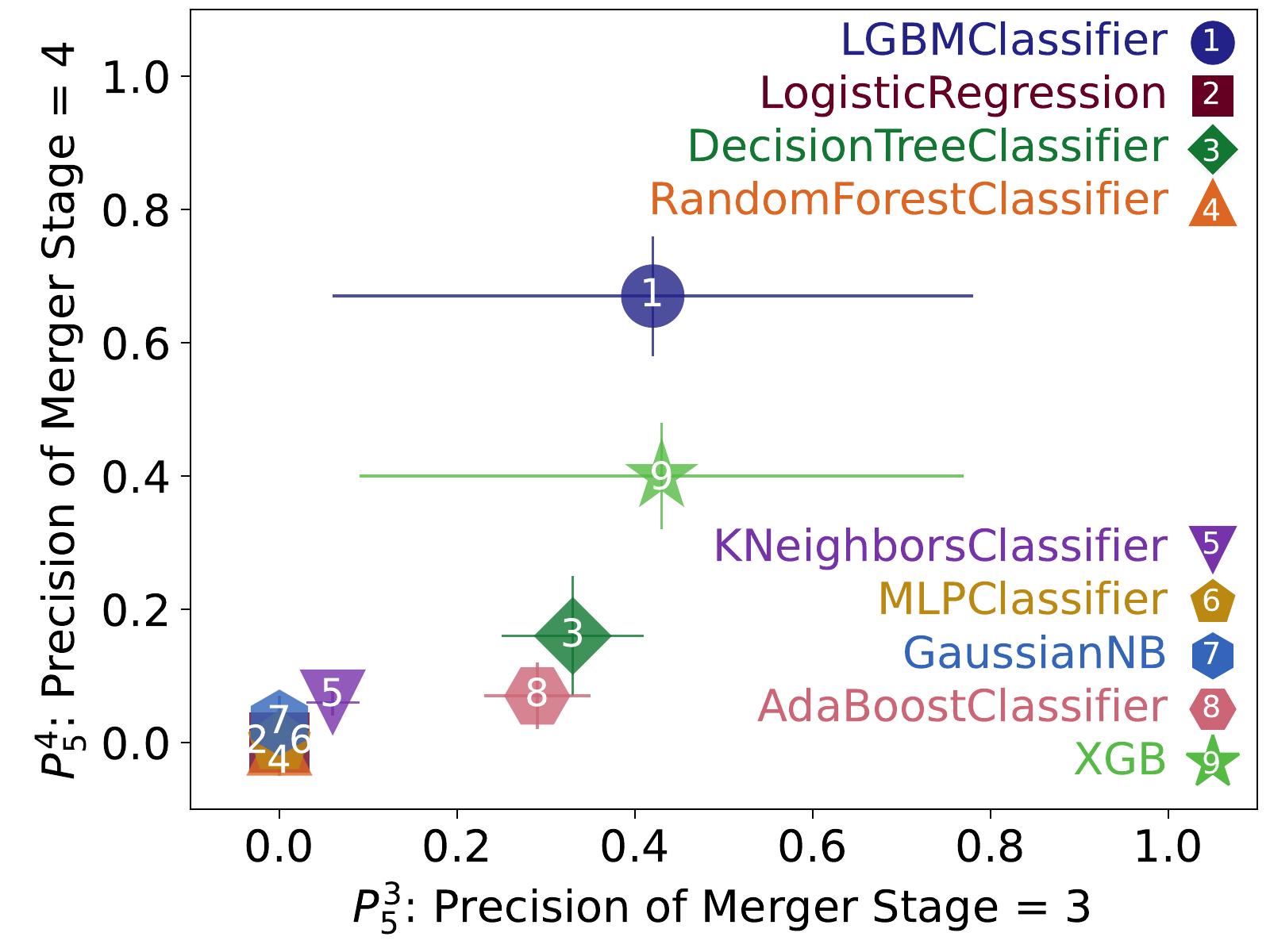} 
\caption{The precision of merger stage 3 and 4 of 5-phase classification for different classifiers with the original sample (N=4,690). }
\label{classifiers534}
\end{figure}

We split the galaxy sample to two-third of them as training set and one-third of them as testing set for all classifiers in all cases. We investigate the errors by bootstrapping the sample for each algorithm and estimate their uncertainties. To achieve high performance, we expect to have high accuracy, high precision, high recall, and high F1 score. For galaxy merger stages, we should avoid contamination from non-mergers and wrong classifications. Therefore, the most important goal is to achieve high purity \citep[e.g,][]{2022MNRAS.511..100B}, which corresponds to high precision score. For multi-classification, we should also check individual precision for each merger stages, that is, merger sources in a specific merger stage which are classified as the correct merger stage over that classified merger stage.  For the 5-phase classification, we label $P_5^1$, $P_5^2$, $P_5^3$, and $P_5^4$ for the individual precision of merger stage 1, 2, 3, and 4, as well as $P_5^0$ for non-merger precision. For the 3-phase classification, we label $P_3^1$ and $P_3^2$ for the individual precision of merger stage 1+3, and 2+4, as well as $P_3^0$ for non-merger. For the 2-phase classification, we label $P_2^1$ for mergers, as well as $P_2^0$ for non-mergers. In Figure~\ref{classifiers2} and Figure~\ref{classifiers3}, we compare the precision of non-mergers and mergers of 2-phase and 3-phase classification for different classifiers with the original sample (N=4,690). In Figure~\ref{classifiers512} and Figure~\ref{classifiers534}, we compare the precision of merger stage 1 and 2, as well as merger stage 3 and 4 of 5-phase classification for different classifiers with the original sample (N=4,690).

\begin{figure}
\centering
\includegraphics[width=1.00\columnwidth]{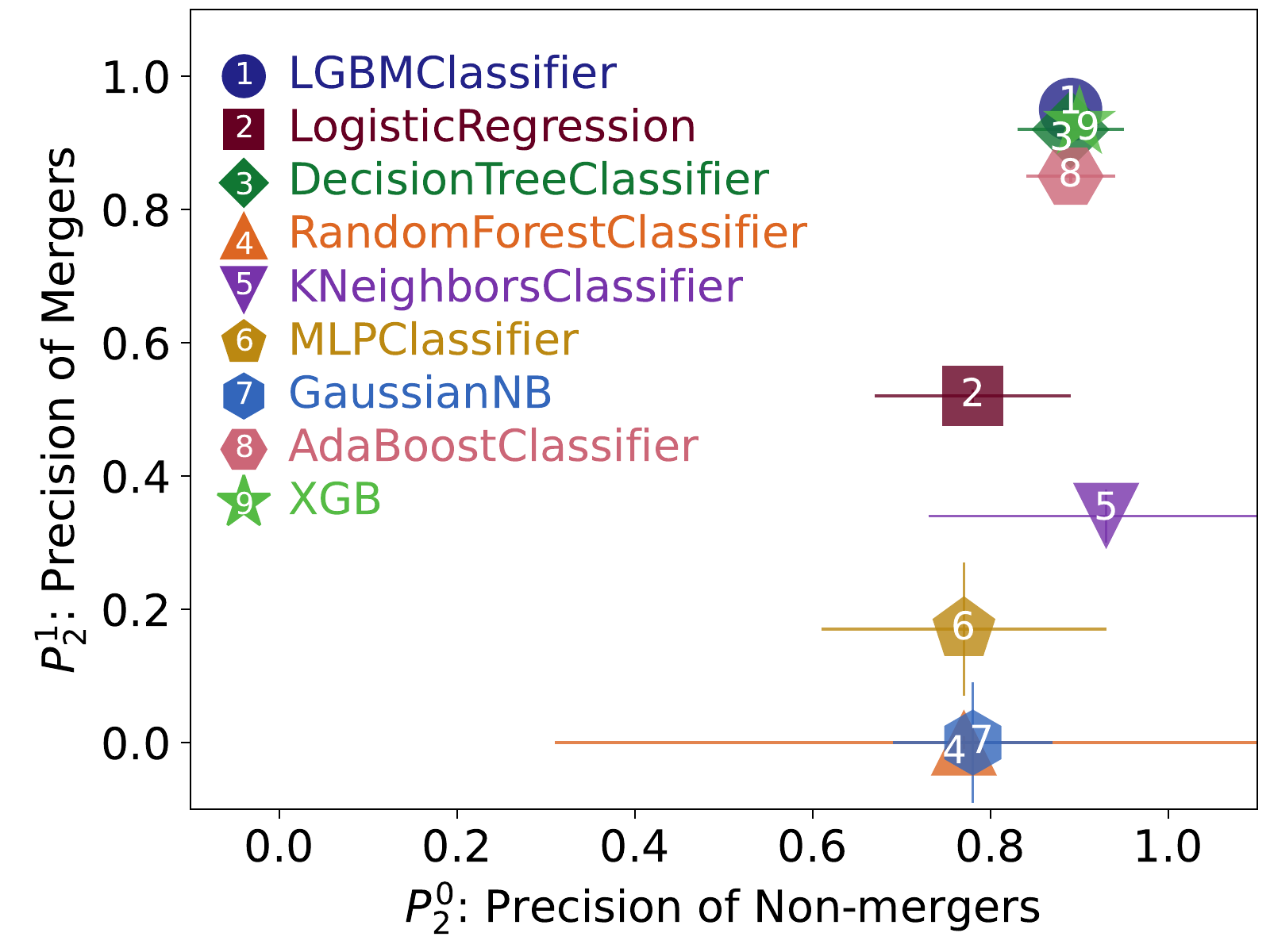} 
\caption{The precision of non-mergers and mergers of 2-phase classificationfor different classifiers for the original with the rotated sample (N=9,380).}
\label{classifiers2_rot}
\end{figure}

\begin{figure}
\centering
\includegraphics[width=1.00\columnwidth]{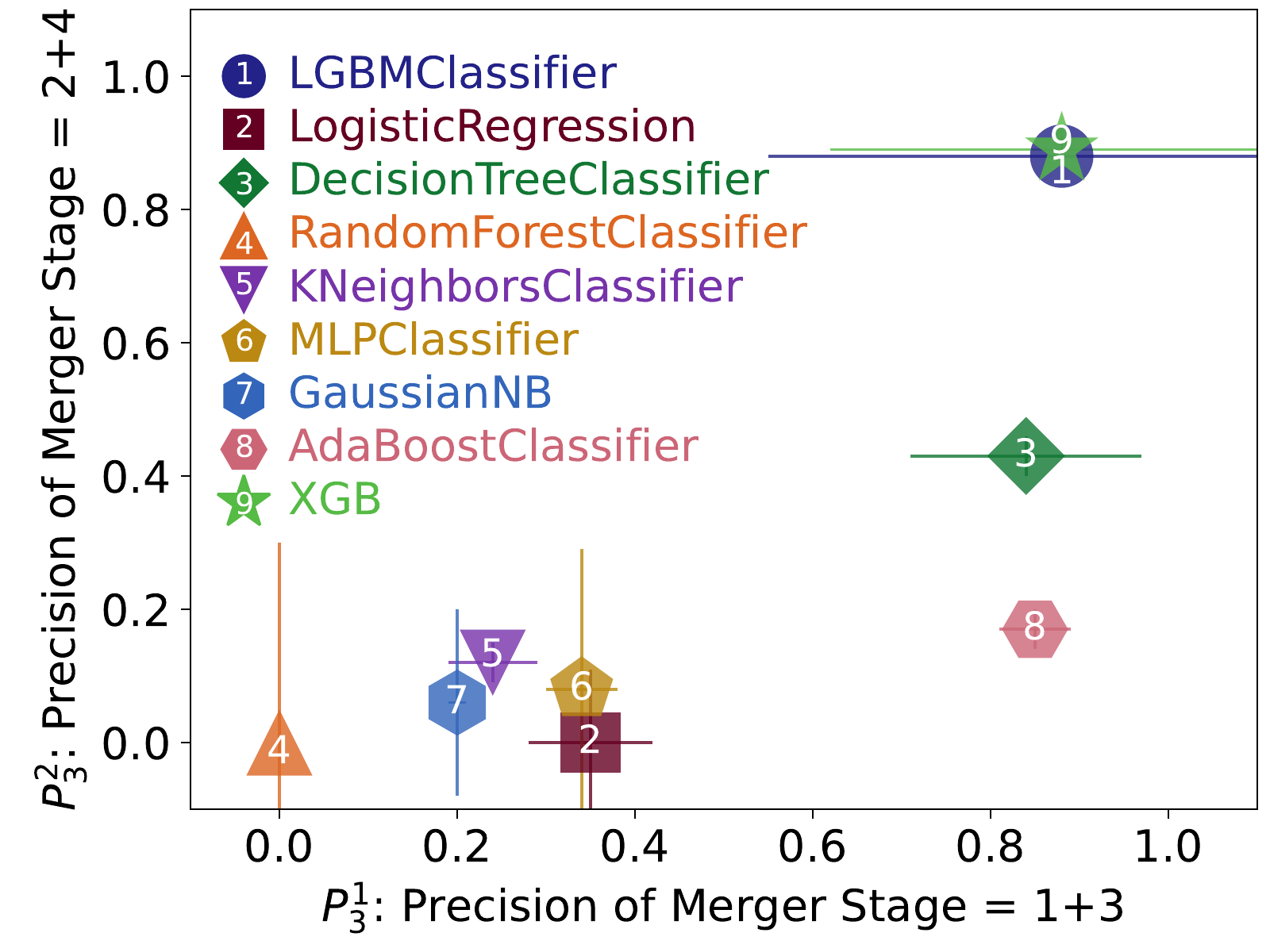} 
\caption{The precision of merger stage 1+3 and 2+4 of 3-phase classificationfor different classifiers for the original with the rotated sample (N=9,380).}
\label{classifiers3_rot}
\end{figure}

\begin{figure}
\centering
\includegraphics[width=1.00\columnwidth]{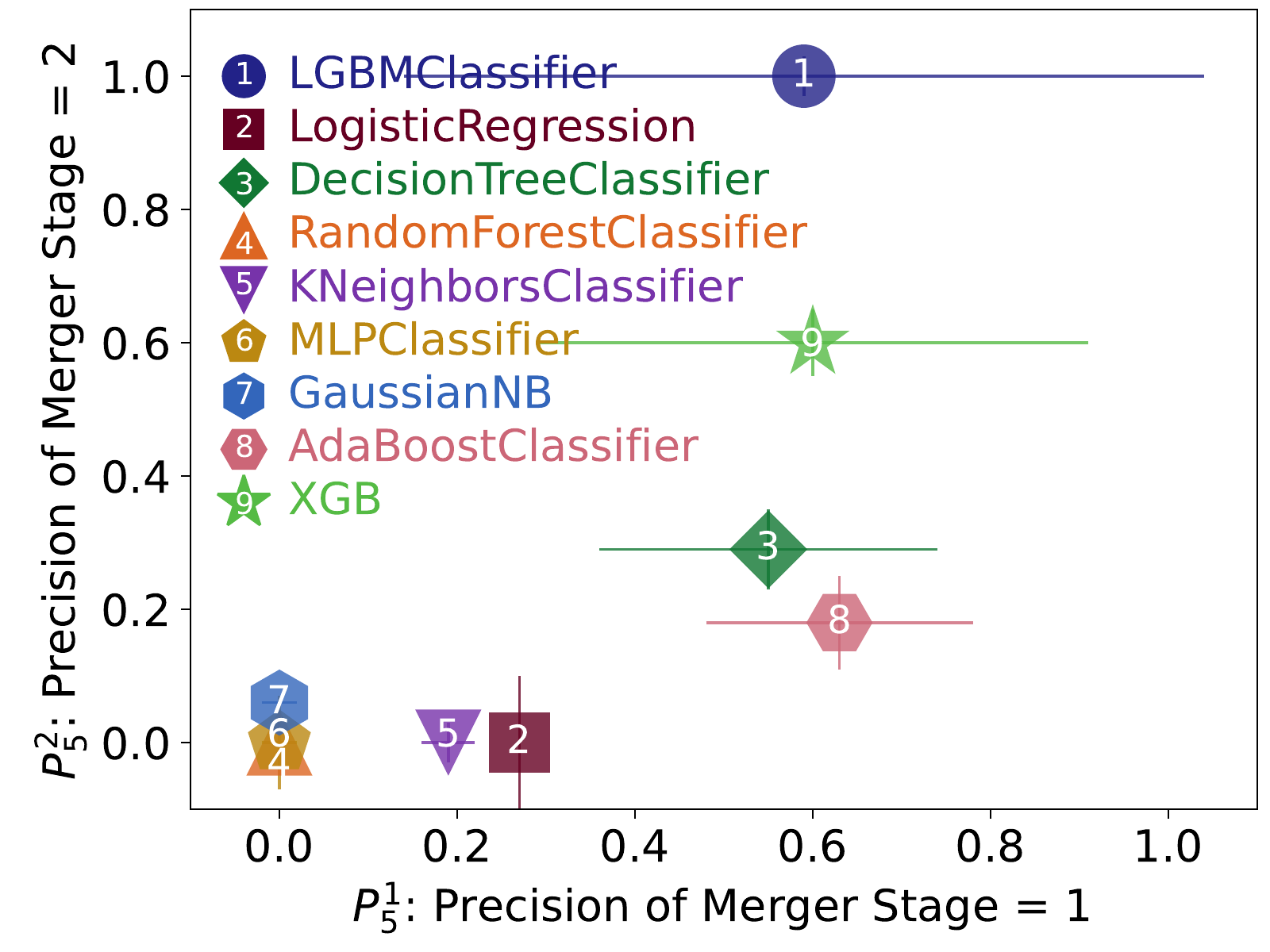} 
\caption{The precision of merger stage 1 and 2 of 5-phase\ classification for different classifiers for the original with the rotated sample (N=9,380).}
\label{classifiers512_rot}
\end{figure}

\begin{figure}
\centering
\includegraphics[width=1.00\columnwidth]{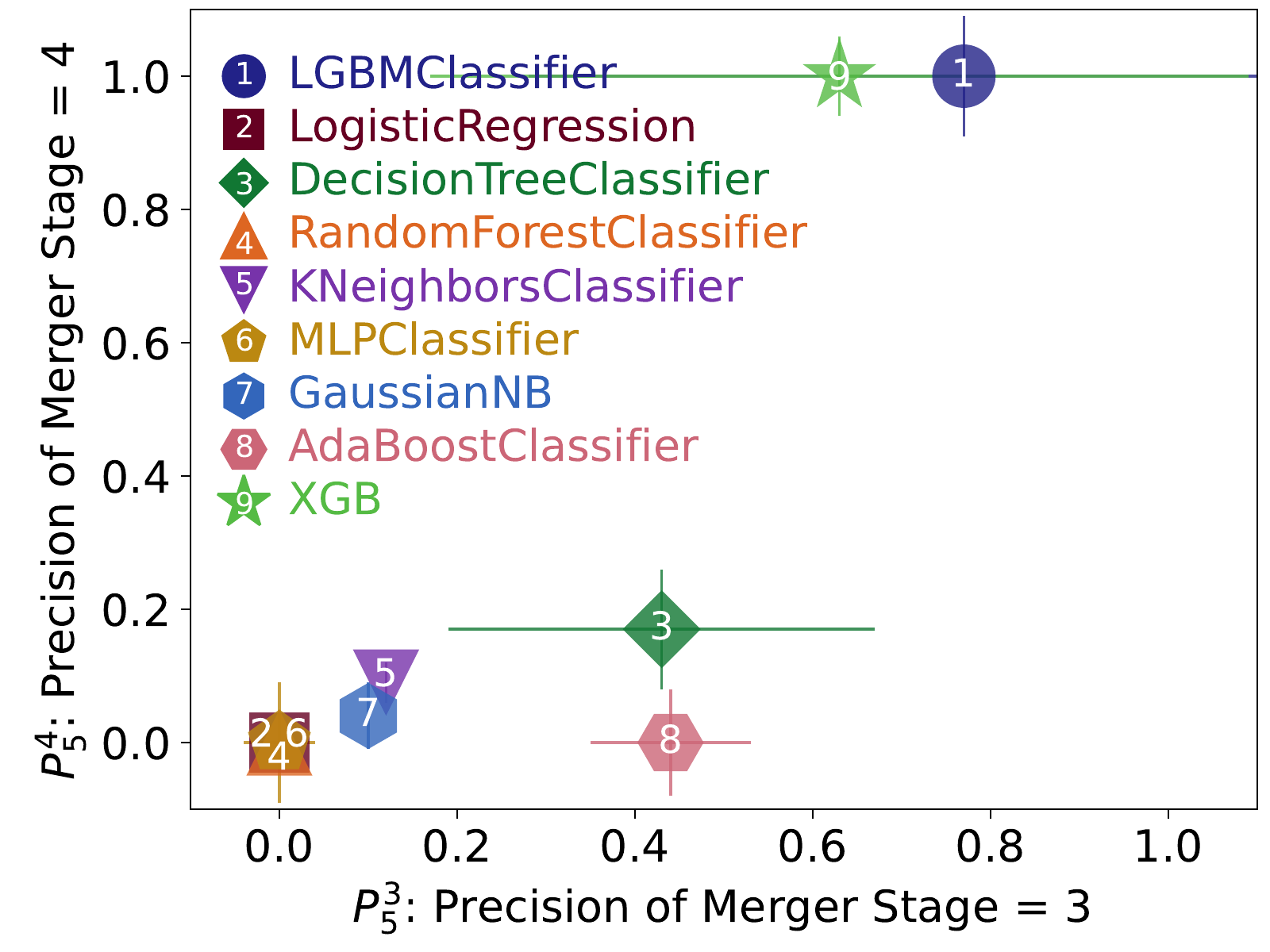} 
\caption{The precision of merger stage 3 and 4 of 5-phase classification for different classifiers for the original with the rotated sample (N=9,380).}
\label{classifiers534_rot}
\end{figure}

For the original SDSS $gri$ images and MaNGA H$\alpha$ velocity maps (N=4,690), we rotate them by 0 and 90 degrees, and we are able to increase their sample size by twice (N=9,380).  In this case, we split the training and testing data before rotation to ensure that an image and its rotated counterpart appear in either the same training or the same test sets. In other words, if an image is used for the training set, its rotation is also only used in the training set.  As a result,  the performance can be improved as shown in Table~\ref{tab1}. The precision can be up to 0.85 for 5-phase classification with \texttt{LGBMClassifier}, but have no significant improvement for 3-phase\ and 2-phase classifications.  We also test the sample by more rotated and flipped combinations, but there are no significant changes. Therefore, we keep the combination of the original and one rotated image (0 and 90 degrees). We show the precision of merger stages of 2-phase, 3-phase, and 5-phase classification for different classifiers with the original sample (N=4,690) in Figure~\ref{classifiers2}, Figure~\ref{classifiers3}, Figure~\ref{classifiers512}, and Figure~\ref{classifiers534}, as well as the original with rotated sample (N=9,380) in Figure~\ref{classifiers2_rot}, Figure~\ref{classifiers3_rot}, Figure~\ref{classifiers512_rot}, and Figure~\ref{classifiers534_rot}.  As a result, the precisions are improved, especially for 5-phase classification with \texttt{LGBMClassifier}.  The 3-phase and 2-phase classifications are slightly improved, but not much as the 5-phase classification as shown in Table~\ref{tab1}. To have a consistent comparison,  we adopt the original with rotated sample (N=9,380)  in the following discussions.

\begin{figure}
\centering
\includegraphics[width=1.00\columnwidth]{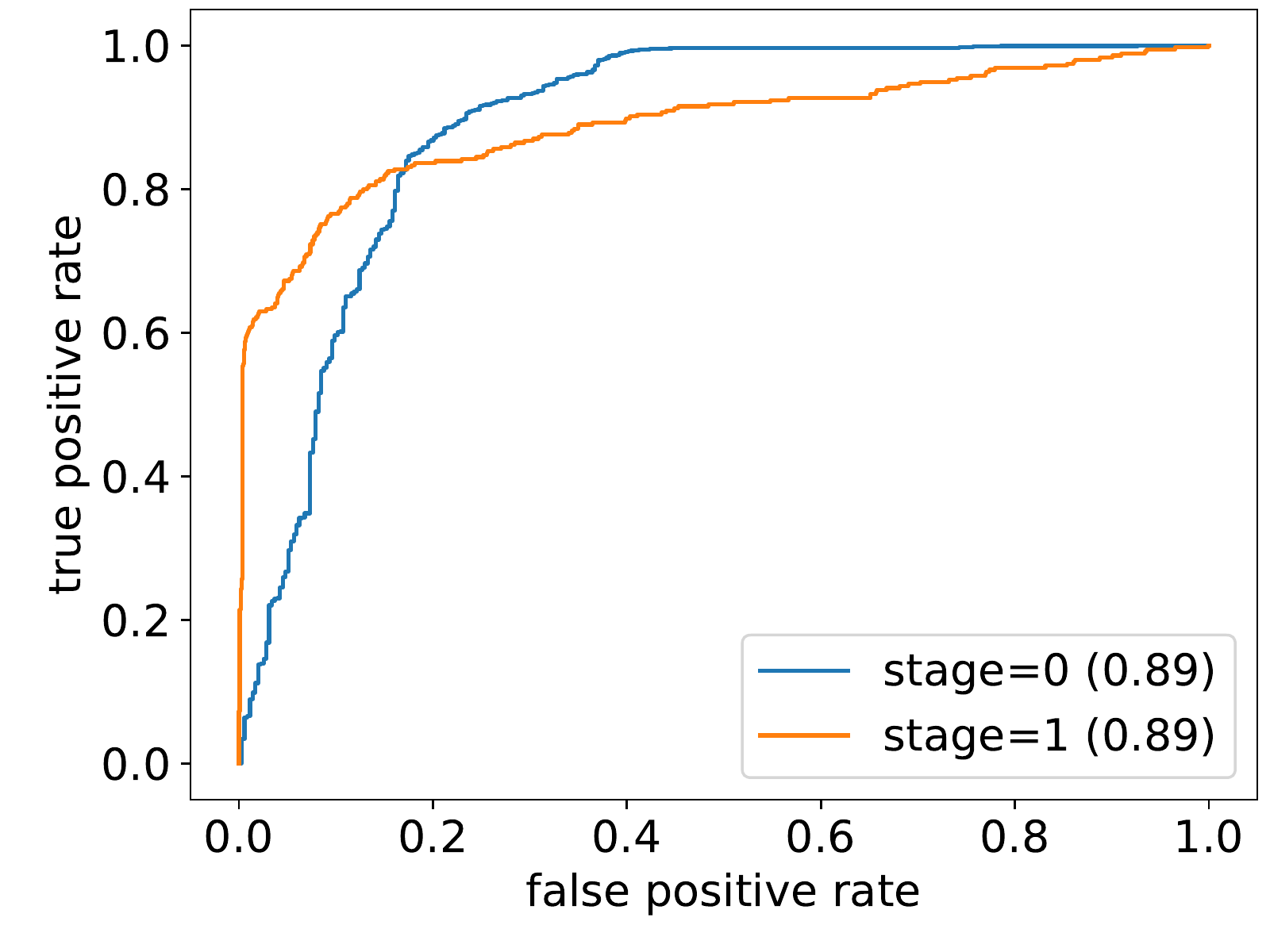} 
\caption{The Receiver Operating Characteristic Curve (ROC) for 2-phase classification. The values show the Area Under the Receiver Operating Characteristic Curve score (AUROC) for the original with rotated sample (N=9,380) with \texttt{LGBMClassifier}. }
\label{roc2}
\end{figure}

\begin{figure}
\centering
\includegraphics[width=1.00\columnwidth]{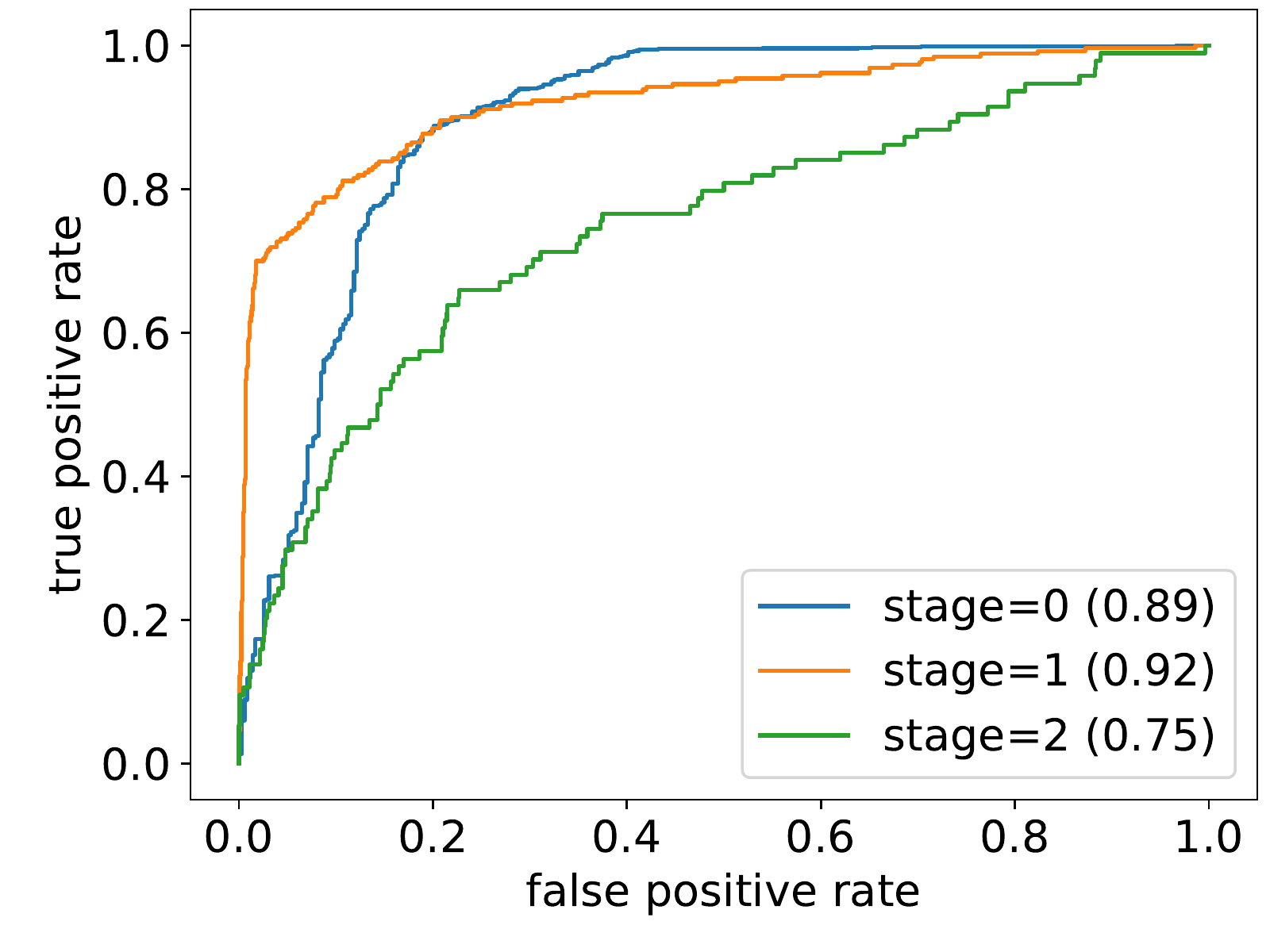} 
\caption{The Receiver Operating Characteristic Curve (ROC) for 3-phase classification. The values show the Area Under the Receiver Operating Characteristic Curve score (AUROC) for the original with rotated sample (N=9,380) with \texttt{LGBMClassifier}. }
\label{roc3}
\end{figure}

\begin{figure}
\centering
\includegraphics[width=1.00\columnwidth]{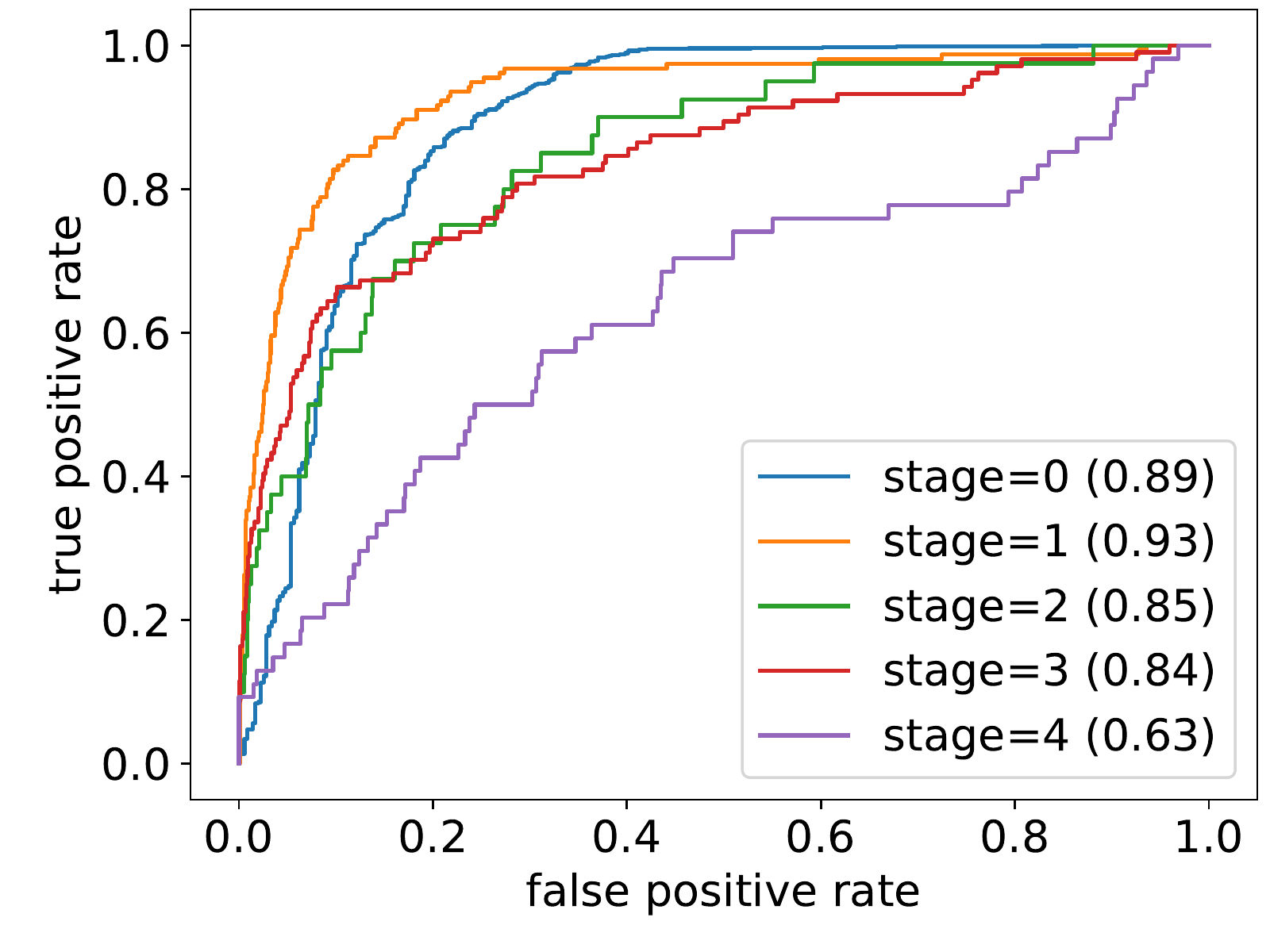} 
\caption{The Receiver Operating Characteristic Curve (ROC) for 5-phase classification. The values show the Area Under the Receiver Operating Characteristic Curve score (AUROC) for the original with rotated sample (N=9,380) with \texttt{LGBMClassifier}. }
\label{roc5}
\end{figure}

Figure~\ref{roc2}, Figure~\ref{roc3}, and Figure~\ref{roc5} shows the Receiver Operating Characteristic Curve (ROC) for the best results (N=9,380) of different phase classification with \texttt{LGBMClassifier}. This is to illustrate the diagnostic ability of the classifier system by plotting the true positive rate (also known as recall or sensitivity) against the false positive rate (also known as probability of false alarm) at various threshold settings.  For multiclass problems, ROC curves and Area Under the ROC scores (AUROC; ideal case is equal to 1) represent each class versus the rest individually. Here we show that the scores can reach high performance ($\gtrsim$0.80) for most cases. 

\begin{figure}
\centering
\includegraphics[width=1.00\columnwidth]{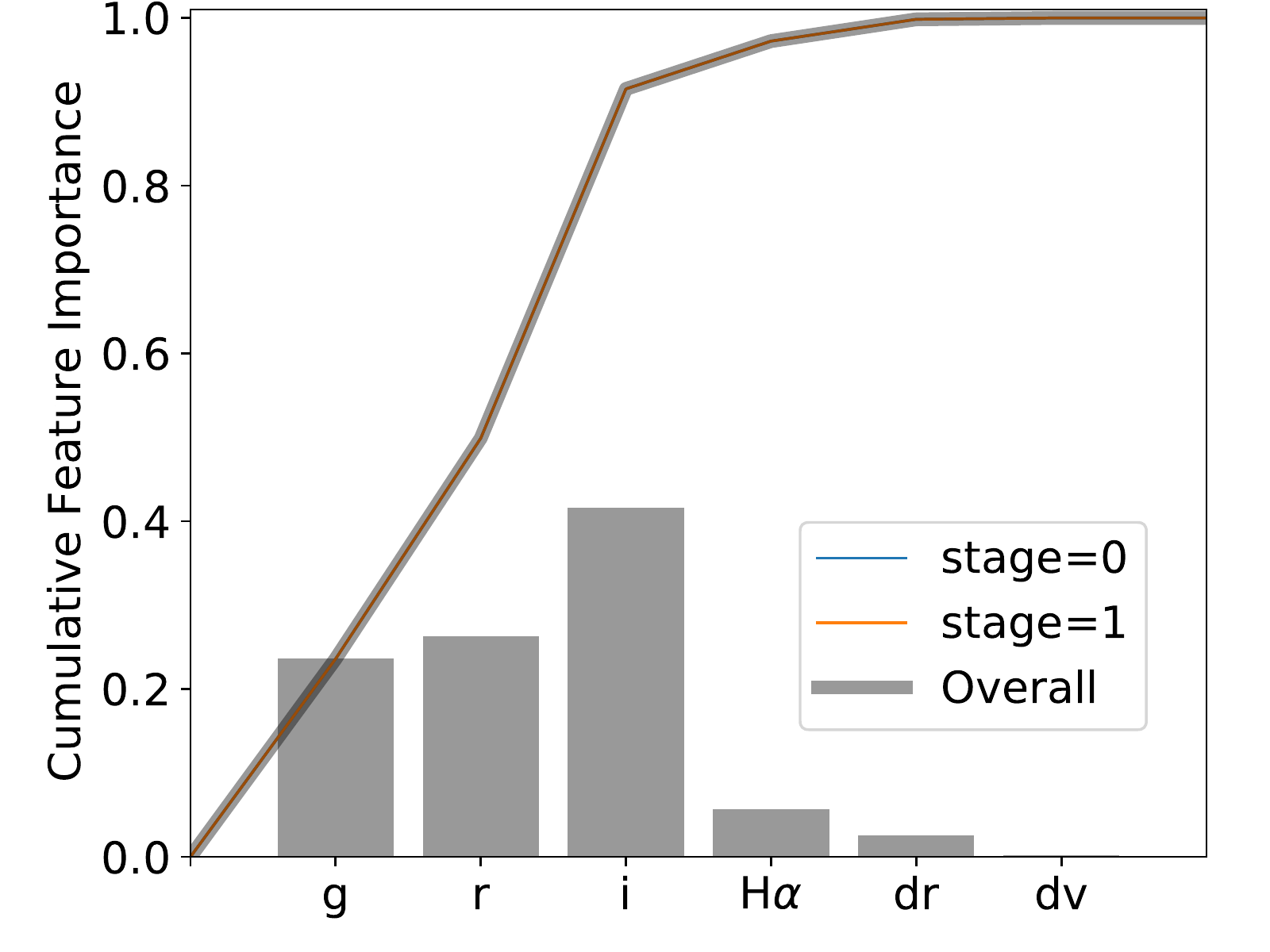} 
\caption{Cumulative feature importance plot for the original with the rotated sample (N=9,380) with \texttt{LGBMClassifier} for individual stages and overall classifications of 2-phase classification. }
\label{importance2}
\end{figure}

\begin{figure}
\centering
\includegraphics[width=1.00\columnwidth]{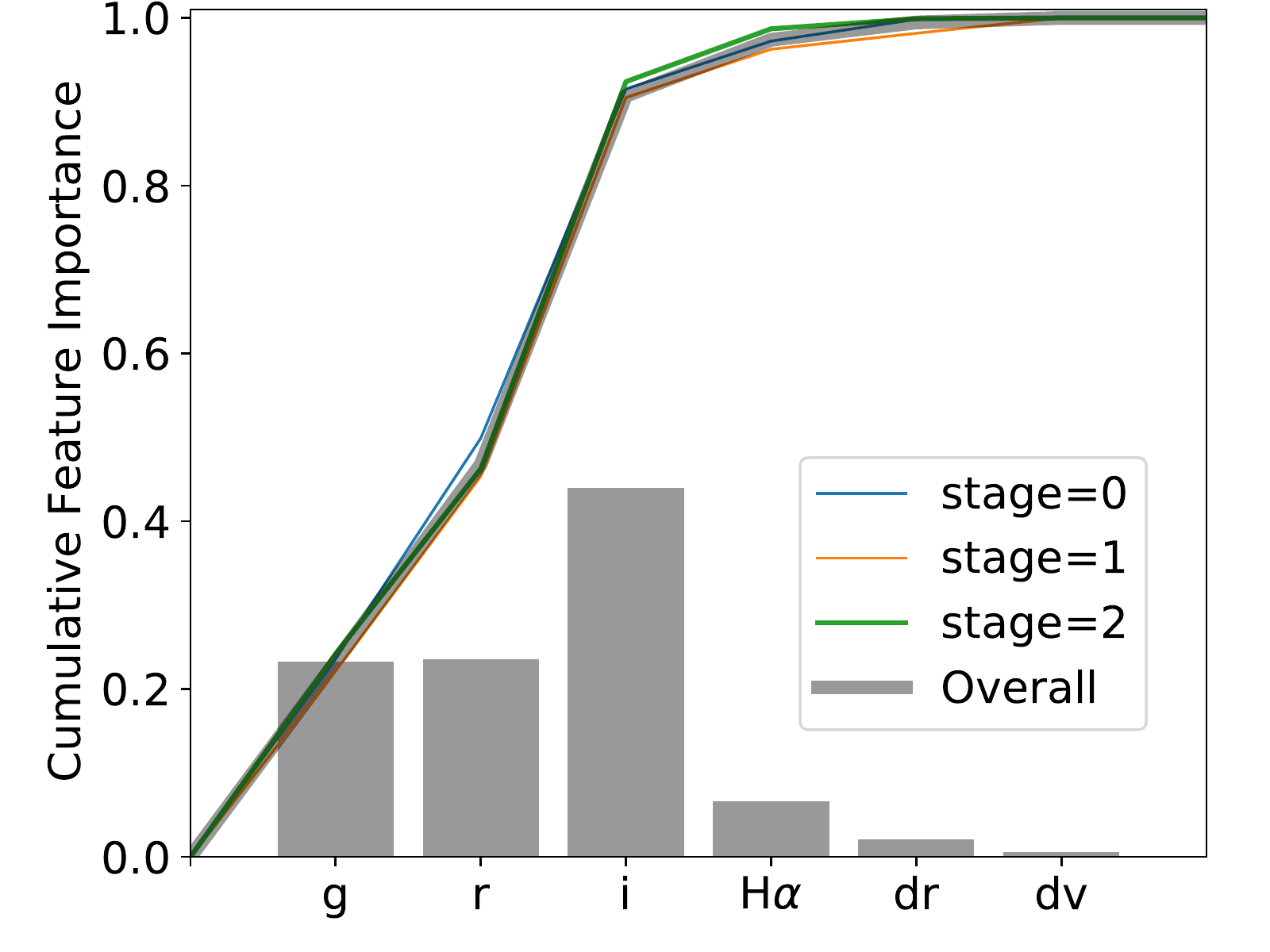} 
\caption{Cumulative feature importance plot for the original with the rotated sample (N=9,380) with \texttt{LGBMClassifier} for individual stages and overall classifications of 3-phase classification. }
\label{importance3}
\end{figure}

\begin{figure}
\centering
\includegraphics[width=1.00\columnwidth]{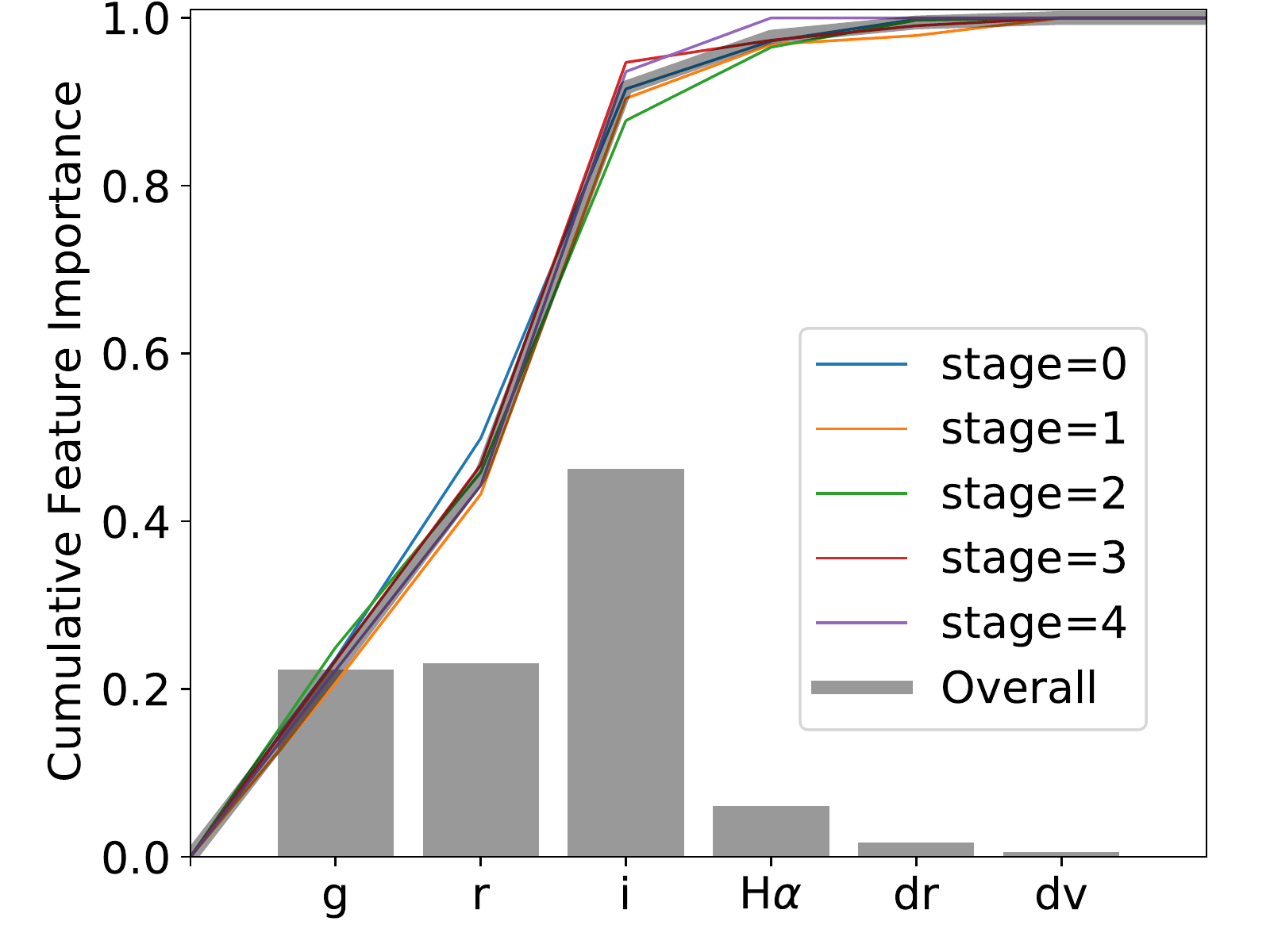} 
\caption{Cumulative feature importance plot for the original with the rotated sample (N=9,380) with \texttt{LGBMClassifier} for individual stages and overall classifications of 5-phase classification. }
\label{importance5}
\end{figure}

We show the feature importance to indicate the importance of each input feature in Figure~\ref{importance2}, Figure~\ref{importance3}, and Figure~\ref{importance5} for the best results (N=9,380) with \texttt{LGBMClassifier}.  In the input parameters of \texttt{LGBMClassifier}, we choose `split' importance type, which calculates numbers of times the feature is used in the tree-structured nodes for each feature. For SDSS $gri$ images and MaNGA H$\alpha$ velocity map, the number of times in the nodes are calculated for each spaxel.  We find that the spaxels in the center are more important than the outskirts, but the results are similar if we rotate, flip, or translate the image by few pixels. To investigate the effect of the entire image size, we sum up the total numbers of times in the tree-structured nodes for all spaxels. Therefore, there are six features: SDSS $g$-band image ($g$),  SDSS $r$-band image ($r$), SDSS $i$-band image($i$), MaNGA H$\alpha$ velocity map (H$\alpha$), the projected separation ($dr$), and the velocity difference ($dv$).  As a sanitary test, we resample the images ($g$, $r$, $i$, and H$\alpha$) to the same scale (e.g., 100 pixels$\times$100 pixels), as well as duplicate the values of $dr$ and $dv$ to the same number of scale (e.g, 100$\times$100). As a result, we find that the ranking of the six features is unchanged, and there are no significant differences between the original and the resampled feature importance.  This indicates that the image size is independent of the feature importance, and the summation of the total numbers of times with all spaxels for each single image is feasible. 

As shown in Figure~\ref{importance2}, Figure~\ref{importance3}, and Figure~\ref{importance5}, the top three important features are SDSS $gri$ images. Among them, SDSS $i$-band image is the most important feature for all 2-, 3-, and 5-phase classifications. The fourth important feature is H$\alpha$, the fifth important feature is $dr$ , the sixth important feature is $dv$. The differences among different stages, which calculated the cumulative feature importance for each stage individually, are not very significant. This is consistent with our expectation because the merger stages were mainly classified by the SDSS imaging. 

\section{Discussion} \label{sec4}

\subsection{Can galaxy interactions be identified?}
\label{sec41}

In this work we use \texttt{LGBMClassifier}, \texttt{LogisticRegression}, \texttt{DecisionTreeClassifier}, \texttt{RandomForestClassifier}, \texttt{KNeighborsClassifier}, \texttt{MLPClassifier}, \texttt{AdaBoostClassifier},  \texttt{GaussianNB},  and \texttt{XGBoost} algorithms to identify interacting galaxies. 

For the 2-phase classification, it is not difficult for some classifiers to reach high accuracy (ACC), high precision (P), high recall (R), and high F1 (F1) score as shown in Table~\ref{tab1}.  We can reach high performance with 0.91 in accuracy, 0.93 in precision (purity), 0.81 in recall (completeness), and 0.85 in F1 score. The good performance of binary classification for merging galaxies is also discussed in \citet[0.53-0.92 in accuracy]{2019A&A...626A..49P}, \citet[0.90 in accuracy]{2020ApJ...895..115F}, \citet[0.80 in accuracy and 0.90 in precision]{2021ApJ...912...45N}, and \citet[0.93 in purity and 0.93 in completeness based on noiseless data from simulations]{2022MNRAS.511..100B}.  The most important score in 2-phase classification is the $P_2^1$ value, which is derived by the merger sources which are classified correctly as mergers over all classified mergers. There are several classifiers (\texttt{LGBMClassifier}, \texttt{XGBoost},  \texttt{DecisionTreeClassifier},  and \texttt{AdaBoostClassifier}) can achieve high $P_2^1$ values ($\gtrsim$0.80) with the original sources (N=4,690) as shown in Figure~\ref{classifiers2}.  In particular, \texttt{LGBMClassifier} and \texttt{XGBoost} can achieve very high $P_2^1$ values ($\gtrsim$0.95).  This may suggest that merger features of interacting galaxies are easier to be identified by tree-structured classifiers than others, especially with boosting trees. 

From the 4,690 galaxies, we select 3,303 star-forming galaxies with $\log$(sSFR/$\rm{yr}^{-1}$)$>$-11, and find that the precision can be improved by 5 to 10 \% even though the sample size is slightly decreased.  One reason might be that  MaNGA H$\alpha$ velocity maps are mainly available for star-forming galaxies or galaxies above some gas fraction, and are blank or low values for gas-poor galaxies which lack spaxels with sufficient signal-to-noise ratio. Nevertheless, the contribution of H$\alpha$ velocity maps is less significant than the SDSS $gri$-band image as shown in Figure~\ref{importance2} and Figure~\ref{data2}. Another reason is that our classification scheme is not easy to distinguish the 5-phase mergers for dry mergers \citep[e.g., ][]{2010MNRAS.404..590L}, which shows fewer distortions on the images anyway.  On the other hand, this would not be a problem for 2-phase or 3-phase mergers, so the differences are less significant. Therefore, visual classification for wet mergers is more reliable than dry mergers, and interaction features are more obvious for star-forming galaxies.This can be consistent with previous finding that classification performance based on both imaging and stellar kinematics tends to increase with high gas fraction \citep{2011ApJ...742..103L,2022MNRAS.511..100B}.

\subsection{Can galaxy merger stages be classified?}

For the 5-phase classification with original sample (N=4,690), the accuracy (ACC) and the precision of non-mergers $P_5^0$ can reach to high scores ($\gtrsim$0.80 and $\gtrsim$0.90, respectively) for several classifiers (\texttt{LGBMClassifier},  \texttt{DecisionTreeClassifier}, \texttt{AdaBoostClassifier}, and \texttt{XGBoost}). If we consider the Top-N accuracy, it can be up to 0.85, 0.92, 9.95, 0.98, 1.00 for N=1 to N=5 cases. However,  the averaged precision (P) and the precision of each stage ($P_5^0$, $P_5^1$, $P_5^2$, $P_5^3$, and $P_5^4$) are less than 0.60 for all classifiers, perhaps because of the degeneracy of the stages which are not easy to be classified by the original sample size.

We combine similar morphology to the same classification in the 3-phase classification, which contains non-merger, 1+3 merger stage for well-separated sources, and 2+4 merger stage for very close pairs.  Table~\ref{tab1} shows that the precision of merger stage 1+3 ($P_3^1$) and merger stage 2+4 ($P_3^2$) can be improved. For \texttt{XGBoost} and \texttt{LGBMClassifier} classifiers, the precision of non-mergers and stage 1+3 can reach high scores ($\gtrsim$0.85), and the precision of non-mergers and stage 2+4 is also improved ($\gtrsim$0.60) as shown in Figure~\ref{classifiers3}.

In order to improve performance of the 5-phase classification, we adopt the combination of original and rotated images. In this work, the physical parameters,such as stellar mass, star formation rate, and interactions between galaxies, would not be affected if we rotate or flip the images. Therefore, we are able to increase the sample size by twice (N=9,380) as shown in Table~\ref{tab1}. We find that the performance of the scores are improved, especially for \texttt{LGBMClassifier} classifiers. In Figure~\ref{classifiers512_rot} and Figure~\ref{classifiers534_rot}, the averaged precision (P) and the precision of non-mergers, stage 2 and stage 4 ($P_5^0$, $P_5^2$, and $P_5^4$) can reach high scores ($\gtrsim$0.90), and the precision of stage 1 and stage 3 ($P_5^1$ and $P_5^3$) is also improved ($\gtrsim$0.60).

We find that \texttt{LGBMClassifier} can provide better performances, especially for good precision, among all classifiers in this work. We have tested different input parameters of \texttt{LGBMClassifier}, but there is no significant improvement for the performance. In some cases, \texttt{XGBoost}, \texttt{DecisionTreeClassifier}, and \texttt{AdaBoostClassifier} can also provide good performances. We test the classifiers by tuning their input parameters, and the above tree-structured classifiers show better performances in most cases. One reason might be that the whole spaxels of the image data are not the best hyperparameters for other models. Nevertheless, we show that the physical features of interacting galaxies are easier to be classified by tree-structured classifiers, especially with gradient boosting trees such as \texttt{XGBoost} and \texttt{LGBMClassifier}. In general, the differences can be explained by the limit of the algorithm, the choice of our input parameters, and the characteristic of our data.  While it may be possible to find a more sophisticate algorithms to improve the performance, our results show that we are able to classify galaxy mergers with good performance by using the ML techniques tested in this work. 

\subsection{What input data and features are important in galaxy interaction?}

We show the cumulative feature importance of the input data in Figure ~\ref{importance2}, Figure ~\ref{importance3}, and Figure ~\ref{importance5}. The top three important features are SDSS images ($i>g\simeq r$). The contribution from MaNGA H$\alpha$ velocity map (H$\alpha$),  the projected separation ($dr$), and line-of-sight velocity difference ($dv$) can also improve the performance.  As discussed in Section~\ref{sec41}, MaNGA H$\alpha$ velocity maps (H$\alpha$) are mainly available for star-forming galaxies or galaxies above some gas fraction. Moreover, $dr$ and $dv$ values are only available when the redshift of two members (pairs) are measured.  It is possible that SDSS images are top features because they are the only data that all the galaxies have. However, if we limit the sample size to objects with available H$\alpha$ velocity maps, or star forming galaxies with $\log$(sSFR/$\rm{yr}^{-1}$)$>$-11 (N=3,303; high signal-to-noise ratio of H$\alpha$),  or the sample with the projected separation and line-of-sight velocity difference (N=651; available $dr$ and $dv$), the feature importance plots are only slightly different, and the top features are still SDSS images ($g$, $r$, and $i$). 
We also find that the differences of cumulative feature importance among various stages are not significant, so the overall curve is sufficient to describe the features. This result is consistent with our expectation, that is, SDSS images are the dominant features to distinguish galaxy interactions. 

\begin{figure}
\centering
\includegraphics[width=1.00\columnwidth]{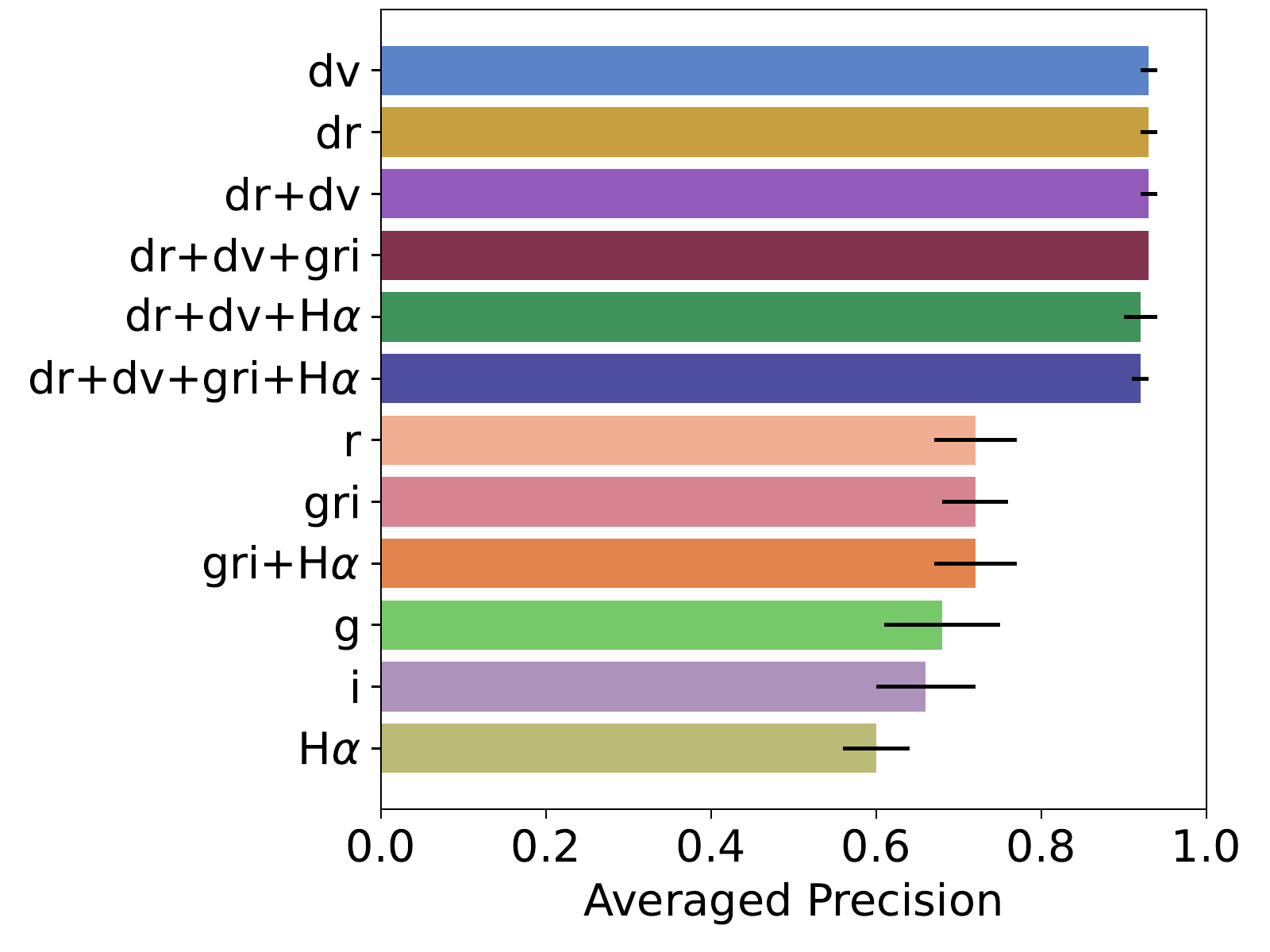} 
\caption{The averaged precision of  2-phase classification for different input data for the original with the rotated sample (N=9,380).}
\label{data2}
\end{figure}

\begin{figure}
\centering
\includegraphics[width=1.00\columnwidth]{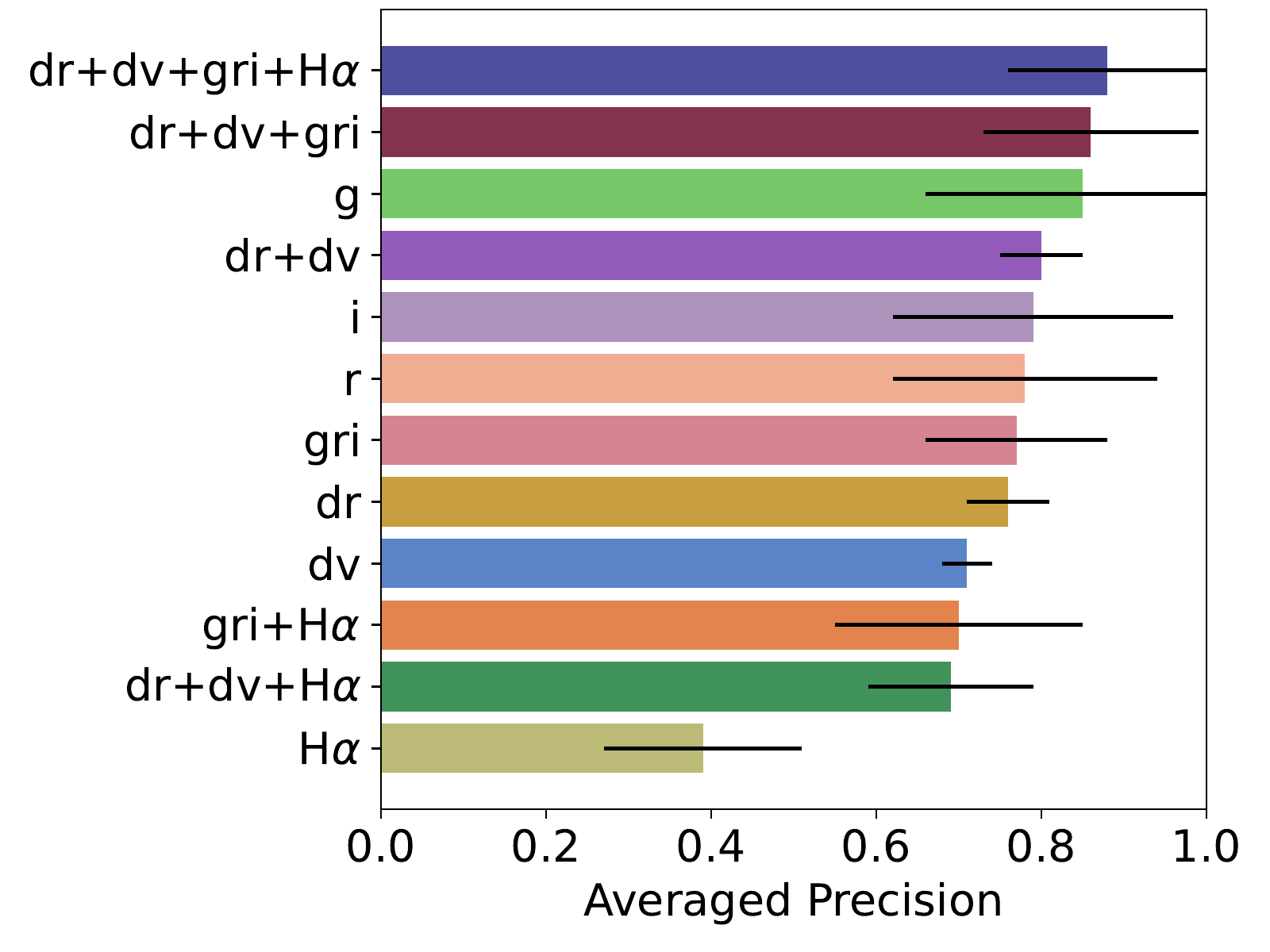} 
\caption{The averaged precision of  3-phase classification for different input data for the original with the rotated sample (N=9,380).}
\label{data3}
\end{figure}

\begin{figure}
\centering
\includegraphics[width=1.00\columnwidth]{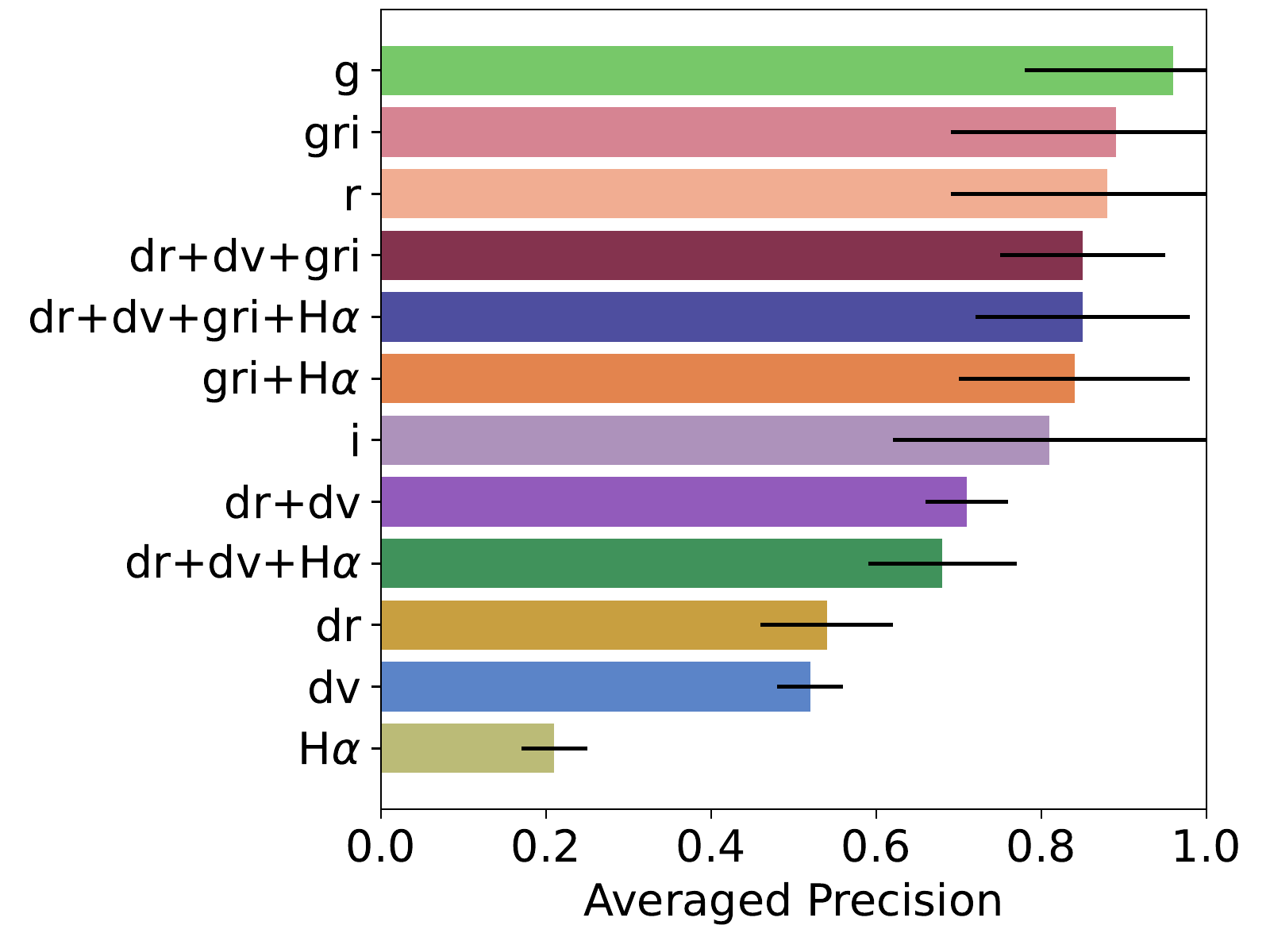} 
\caption{The averaged precision of  5-phase classification for different input data for the original with the rotated sample (N=9,380).}
\label{data5}
\end{figure}

In order to investigate the input features individually, we test the performance of different combinations of input data (the projected separation,  the velocity difference, the SDSS $gri$-band image, and the MaNGA H$\alpha$ velocity map) for original with the rotated sample (N=9,380) by \texttt{LGBMClassifier} in Table~\ref{tab2}, Figure~\ref{data2}, Figure~\ref{data3}, and Figure~\ref{data5}.  In general, the most important features are SDSS $gri$ images,  and the contribution from MaNGA H$\alpha$ velocity map,  the projected separation, as well as line-of-sight velocity difference can also improve the performance by 0-20\%. Because $dr$ and $dv$ only exist for galaxies with spectroscopic neighbors, so they should alone yield all galaxies in the pair phase wherever both galaxies have spectra.  If we remove the information of $dr$ and $dv$, the performance is decreased, especially for 2-phase classification. Moreover,  $dr$ or $dv$ alone can reach high performance for 2-phase classification, but the change in performance is not for 3-phase and 5-phase classifications. This suggests that $dr$ and $dv$ are sufficient to identify galaxy mergers, but additional imaging data are required if detailed merging stages are investigated. For 3-phase classification, the performance is less affected if the SDSS $gri$-band images are still included.  The $g$-, $r$-, and $i$-band images are highly correlated to each other. If we remove one of them, the performance is slightly decreased but not significant compared to the case which we include all of the three $gri$-band. As a result, we show that the contributions from SDSS images are important. For 2-phase classification in Figure~\ref{data2}, $dr$ or $dv$ alone can already achieve a high precision score ($\gtrsim$0.90).  This can be explained by that these two values are only available when redshift of pairs measured, so the information is already sufficient to identify galaxy mergers.  However, for 3-phase classification in Figure~\ref{data3} or 5-phase classification in Figure~\ref{data5}, it is obvious to see that SDSS $gri$-band image and the MaNGA H$\alpha$ velocity map are required. Besides, it is interesting that the information from $g$, $r$, $i$-image alone can each reach $\sim$70\% precision, and even better than some other combinations. This implies that a single-band SDSS image for MaNGA galaxies can provide sufficient photometric quality to reveal the most important information and low-surface brightness features about galaxy interactions \citep[e.g., ][]{2019MNRAS.490.5390B, 2021MNRAS.506..677C, 2021MNRAS.504..372B, 2020ApJ...895..115F}. In particular, SDSS $i$-band image is the top feature in Figure~\ref{importance2},  Figure~\ref{importance3}, and Figure~\ref{importance5}, and single $g$-band SDSS image can achieve a high precision score ($\gtrsim$0.90) for 3-phase and 5-phase classifications as shown in Figure~\ref{data3} and Figure~\ref{data5}.  This implies that single SDSS band image might be sufficient to distinguish detailed interacting features of galaxies, but further investigations are required.

If we remove MaNGA H$\alpha$ velocity map from the input data as shown in Table~\ref{tab2}, the performance might be decreased  but have little changes ($<$5\%) for 2-phase, 3-phase, and 5-phase classifications. \citet{2022MNRAS.511..100B} shows that stellar kinematic data has little to offer in compliment to imaging for merger remnant identification by using TNG100 cosmological hydrodynamical simulation.  Besides, \citet{2021ApJ...912...45N} investigate the importance of stellar kinematics, and find 0.80 in accuracy and 0.90 in precision for major merger classification.  \citet{2022MNRAS.tmp.1734M} use stellar kinematics on the role of field-of-view limitations to classify mergers into different stages, and find 0.40-0.79 completeness in pairs, and 0.97-1.00 completeness in the merging and post-coalescent phases.  In our work, IFU data can still provide redshift information (i.e,  $dv$, the difference in the line-of-sight velocity) to improve the performance of merger stage classification, especially for 2-phase classification.  However, the kinematic information in the velocity map provides little contribution to the current merger stage classification. On the other hand, the image and the velocity information alone can already achieve high performance.  There are many available IFU data from MaNGA, but the current choice of inputs (i. e., projected separation, the velocity difference, all the spaxels of the SDSS $gri$-band image and the MaNGA H$\alpha$ velocity map) are sufficient to identify galaxy merger and their stages with a sample size of 4,690 galaxies with the state-of-the-art classifiers and the rotation technique. It is possible that other measurements in the IFU data can provide more information, such as kinematics of galaxy interactions derived from other MaNGA data. If the dataset is with different quality (i.e., resolution or signal-to-noise ratio), it is also possible that additional IFU data, such as asymmetries and other non-parametric map characteristics,  become important information.  Moreover, we check our false positive sources, which are non-merger galaxies classified by visual inspection while merger galaxies classified by machine learning. There are some suspicious cases, but it is difficult to conclude whether they turn out to be possible interacting galaxies.  Therefore, it is possible that the abundant IFU data from MaNGA may unveil additional information by using unsupervised or other ML techniques.. In this work, we show that the current parameters are already enough to classify galaxy merger stages without visual inspection. This will be helpful for us to understand galaxy interaction with a larger sample in the future all-sky surveys. 


\section{Summary} \label{sec5}

In this paper, we have identified MaNGA merger stages by machine learning techniques. Our main findings are as follows.
\begin{enumerate}

\item For 2-phase classification, the performance can be high (precision$>$0.90) with LGBMClassifier. In general, merger features of interacting galaxies are easy to be identified by tree-structured classifiers, especially with gradient boosting trees. 

\item Sample size can be increased by the combination of the original and rotated images. Then 5-phase classification can also be good (precision$>$0.85).

\item The most important features are SDSS $gri$ images, and single-band SDSS image can already provide most information about galaxy interactions. The contribution from MaNGA H$\alpha$ velocity map,  the projected separation, and line-of-sight velocity difference can also improve the performance by 0-20\%. 

\item The kinematic information in the velocity map provides less contribution to the current merger stage classification. On the other hand, the image and the velocity information alone can already achieve high performance. 

\item These results can apply to the entire MaNGA data as well as future all-sky surveys. 
\end{enumerate}


\begin{table*}[t]
\centering
\scriptsize
\begin{tabular}{cccccccccc}
\hline
\hline
classifier & ACC & P & R & F1 & $P^0$ & $P^1$ & $P^2$ & $P^3$ & $P^4$\\
\hline
\hline
\multicolumn{10}{c}{5-phase Classification (original with rotated sample, N=9,380)} \\ 
\hline
LGBMClassifier & 0.85$\pm$0.01 & 0.85$\pm$0.13 & 0.39$\pm$0.01 & 0.40$\pm$0.02 & 0.89$\pm$0.01 & 0.59$\pm$0.03 & 1.00$\pm$0.45 & 0.77$\pm$0.06 & 1.00$\pm$0.49 \\
LogisticRegression & 0.77$\pm$0.01 & 0.21$\pm$0.02 & 0.20$\pm$0.00 & 0.18$\pm$0.00 & 0.78$\pm$0.01 & 0.27$\pm$0.10 & 0.00$\pm$0.00 & 0.00$\pm$0.00 & 0.00$\pm$0.00 \\
DecisionTreeClassifier & 0.84$\pm$0.02 & 0.47$\pm$0.08 & 0.38$\pm$0.02 & 0.40$\pm$0.03 & 0.90$\pm$0.01 & 0.55$\pm$0.06 & 0.29$\pm$0.19 & 0.43$\pm$0.10 & 0.17$\pm$0.24 \\
RandomForestClassifier & 0.77$\pm$0.01 & 0.15$\pm$0.00 & 0.20$\pm$0.00 & 0.17$\pm$0.00 & 0.77$\pm$0.01 & 0.00$\pm$0.00 & 0.00$\pm$0.00 & 0.00$\pm$0.00 & 0.00$\pm$0.00 \\
KNeighborsClassifier & 0.75$\pm$0.02 & 0.24$\pm$0.01 & 0.21$\pm$0.01 & 0.19$\pm$0.01 & 0.78$\pm$0.01 & 0.19$\pm$0.03 & 0.00$\pm$0.03 & 0.12$\pm$0.03 & 0.09$\pm$0.04 \\
MLPClassifier & 0.77$\pm$0.16 & 0.15$\pm$0.03 & 0.20$\pm$0.02 & 0.17$\pm$0.01 & 0.77$\pm$0.03 & 0.00$\pm$0.07 & 0.00$\pm$0.02 & 0.00$\pm$0.07 & 0.00$\pm$0.00 \\
GaussianNB & 0.06$\pm$0.01 & 0.15$\pm$0.02 & 0.24$\pm$0.02 & 0.06$\pm$0.01 & 0.58$\pm$0.07 & 0.00$\pm$0.00 & 0.06$\pm$0.02 & 0.10$\pm$0.02 & 0.04$\pm$0.01 \\
AdaBoostClassifier & 0.83$\pm$0.01 & 0.43$\pm$0.05 & 0.39$\pm$0.02 & 0.41$\pm$0.03 & 0.89$\pm$0.01 & 0.63$\pm$0.07 & 0.18$\pm$0.15 & 0.44$\pm$0.09 & 0.00$\pm$0.09 \\
XGB  & 0.85$\pm$0.01 & 0.74$\pm$0.13 & 0.41$\pm$0.01 & 0.44$\pm$0.02 & 0.89$\pm$0.01 & 0.60$\pm$0.05 & 0.60$\pm$0.31 & 0.63$\pm$0.10 & 1.00$\pm$0.46 \\
\hline
\multicolumn{10}{c}{5-phase Classification (original sample, N=4,690)} \\ 
\hline
LGBMClassifier & 0.85$\pm$0.01 & 0.72$\pm$0.08 & 0.38$\pm$0.02 & 0.40$\pm$0.02 & 0.90$\pm$0.01 & 0.62$\pm$0.04 & 1.00$\pm$0.33 & 0.42$\pm$0.09 & 0.67$\pm$0.36 \\
LogisticRegression & 0.77$\pm$0.01 & 0.15$\pm$0.03 & 0.20$\pm$0.00 & 0.17$\pm$0.00 & 0.77$\pm$0.01 & 0.00$\pm$0.17 & 0.00$\pm$0.00 & 0.00$\pm$0.00 & 0.00$\pm$0.00 \\
DecisionTreeClassifier & 0.83$\pm$0.00 & 0.41$\pm$0.03 & 0.37$\pm$0.01 & 0.38$\pm$0.02 & 0.90$\pm$0.01 & 0.61$\pm$0.05 & 0.05$\pm$0.08 & 0.33$\pm$0.09 & 0.16$\pm$0.08 \\
RandomForestClassifier & 0.77$\pm$0.01 & 0.15$\pm$0.00 & 0.20$\pm$0.00 & 0.17$\pm$0.00 & 0.77$\pm$0.01 & 0.00$\pm$0.00 & 0.00$\pm$0.00 & 0.00$\pm$0.00 & 0.00$\pm$0.00 \\
KNeighborsClassifier & 0.74$\pm$0.01 & 0.22$\pm$0.02 & 0.20$\pm$0.02 & 0.19$\pm$0.02 & 0.77$\pm$0.01 & 0.21$\pm$0.06 & 0.00$\pm$0.05 & 0.06$\pm$0.02 & 0.06$\pm$0.03 \\
MLPClassifier & 0.77$\pm$0.16 & 0.15$\pm$0.05 & 0.20$\pm$0.02 & 0.17$\pm$0.03 & 0.77$\pm$0.03 & 0.00$\pm$0.27 & 0.00$\pm$0.02 & 0.00$\pm$0.05 & 0.00$\pm$0.01 \\
GaussianNB & 0.08$\pm$0.02 & 0.20$\pm$0.03 & 0.26$\pm$0.03 & 0.06$\pm$0.01 & 0.72$\pm$0.07 & 0.17$\pm$0.07 & 0.06$\pm$0.03 & 0.00$\pm$0.04 & 0.03$\pm$0.01 \\
AdaBoostClassifier & 0.82$\pm$0.01 & 0.37$\pm$0.02 & 0.34$\pm$0.02 & 0.35$\pm$0.02 & 0.90$\pm$0.01 & 0.56$\pm$0.04 & 0.06$\pm$0.06 & 0.29$\pm$0.05 & 0.07$\pm$0.06 \\
XGB & 0.85$\pm$0.01 & 0.55$\pm$0.06 & 0.39$\pm$0.02 & 0.42$\pm$0.02 & 0.90$\pm$0.01 & 0.64$\pm$0.04 & 0.38$\pm$0.15 & 0.43$\pm$0.08 & 0.40$\pm$0.34 \\
\hline
\hline
\multicolumn{8}{c}{3-phase Classification (original with rotated sample, N=9,380)} \\ 
\hline
LGBMClassifier & 0.89$\pm$0.01 & 0.88$\pm$0.12 & 0.59$\pm$0.01 & 0.62$\pm$0.03 & 0.89$\pm$0.01 & 0.88$\pm$0.03 & 0.88$\pm$0.33 & & \\
LogisticRegression & 0.77$\pm$0.01 & 0.38$\pm$0.04 & 0.34$\pm$0.01 & 0.31$\pm$0.01 & 0.78$\pm$0.01 & 0.35$\pm$0.11 & 0.00$\pm$0.07 & & \\
DecisionTreeClassifier & 0.88$\pm$0.01 & 0.72$\pm$0.04 & 0.59$\pm$0.01 & 0.62$\pm$0.02 & 0.89$\pm$0.01 & 0.84$\pm$0.03 & 0.43$\pm$0.13 & & \\
RandomForestClassifier & 0.77$\pm$0.01 & 0.26$\pm$0.10 & 0.33$\pm$0.00 & 0.29$\pm$0.00 & 0.77$\pm$0.01 & 0.00$\pm$0.30 & 0.00$\pm$0.00 & & \\
KNeighborsClassifier & 0.74$\pm$0.01 & 0.38$\pm$0.02 & 0.35$\pm$0.02 & 0.34$\pm$0.02 & 0.78$\pm$0.01 & 0.24$\pm$0.03 & 0.12$\pm$0.05 & & \\
MLPClassifier & 0.65$\pm$0.14 & 0.44$\pm$0.09 & 0.50$\pm$0.06 & 0.44$\pm$0.07 & 0.90$\pm$0.05 & 0.34$\pm$0.21 & 0.08$\pm$0.04 & & \\
GaussianNB & 0.07$\pm$0.01 & 0.29$\pm$0.05 & 0.33$\pm$0.01 & 0.05$\pm$0.01 & 0.61$\pm$0.12 & 0.20$\pm$0.14 & 0.06$\pm$0.01 & & \\
AdaBoostClassifier & 0.87$\pm$0.01 & 0.64$\pm$0.02 & 0.56$\pm$0.01 & 0.59$\pm$0.01 & 0.89$\pm$0.01 & 0.85$\pm$0.03 & 0.17$\pm$0.04 & & \\
XGB & 0.89$\pm$0.01 & 0.89$\pm$0.09 & 0.59$\pm$0.01 & 0.63$\pm$0.01 & 0.89$\pm$0.01 & 0.88$\pm$0.03 & 0.89$\pm$0.26 & & \\
\hline
\multicolumn{8}{c}{3-phase Classification (original sample, N=4,690)} \\ 
\hline
LGBMClassifier & 0.89$\pm$0.01 & 0.79$\pm$0.12 & 0.58$\pm$0.02 & 0.60$\pm$0.02 & 0.90$\pm$0.01 & 0.87$\pm$0.02 & 0.60$\pm$0.35 & & \\
LogisticRegression  & 0.77$\pm$0.01 & 0.33$\pm$0.03 & 0.33$\pm$0.00 & 0.30$\pm$0.00 & 0.77$\pm$0.01 & 0.21$\pm$0.09 & 0.00$\pm$0.00 & & \\
DecisionTreeClassifier & 0.87$\pm$0.01 & 0.65$\pm$0.02 & 0.60$\pm$0.01 & 0.62$\pm$0.02 & 0.91$\pm$0.01 & 0.85$\pm$0.04 & 0.21$\pm$0.05 & & \\
RandomForestClassifier & 0.77$\pm$0.01 & 0.26$\pm$0.13 & 0.33$\pm$0.00 & 0.29$\pm$0.00 & 0.77$\pm$0.01 & 0.00$\pm$0.40 & 0.00$\pm$0.00 & & \\
KNeighborsClassifier &0.74$\pm$0.02 & 0.36$\pm$0.02 & 0.34$\pm$0.02 & 0.33$\pm$0.02 & 0.78$\pm$0.01 & 0.23$\pm$0.03 & 0.07$\pm$0.03 & & \\
MLPClassifier & 0.76$\pm$0.19 & 0.46$\pm$0.06 & 0.39$\pm$0.02 & 0.39$\pm$0.05 & 0.80$\pm$0.04 & 0.42$\pm$0.18 & 0.16$\pm$0.03 & & \\
GaussianNB & 0.08$\pm$0.03 & 0.31$\pm$0.05 & 0.33$\pm$0.01 & 0.06$\pm$0.01 & 0.70$\pm$0.08 & 0.17$\pm$0.09 & 0.05$\pm$0.01 & & \\
AdaBoostClassifier & 0.87$\pm$0.01 & 0.61$\pm$0.03 & 0.55$\pm$0.02 & 0.57$\pm$0.02 & 0.90$\pm$0.01 & 0.85$\pm$0.04 & 0.07$\pm$0.08 & & \\
XGB & 0.89$\pm$0.01 & 0.84$\pm$0.09 & 0.59$\pm$0.01 & 0.62$\pm$0.01 & 0.90$\pm$0.01 & 0.87$\pm$0.02 & 0.75$\pm$0.28 & & \\
\hline
\hline
\multicolumn{7}{c}{2-phase Classification (original with rotated sample, N=9,380)} \\ 
\hline
LGBMClassifier & 0.90$\pm$0.00 & 0.92$\pm$0.01 & 0.79$\pm$0.01 & 0.84$\pm$0.01 & 0.89$\pm$0.01 & 0.95$\pm$0.01 & & & \\
LogisticRegression & 0.77$\pm$0.01 & 0.65$\pm$0.06 & 0.51$\pm$0.01 & 0.47$\pm$0.01 & 0.78$\pm$0.01 & 0.52$\pm$0.11 & & & \\
DecisionTreeClassifier & 0.89$\pm$0.01 & 0.91$\pm$0.03 & 0.78$\pm$0.01 & 0.83$\pm$0.01 & 0.89$\pm$0.01 & 0.92$\pm$0.06 & & & \\
RandomForestClassifier & 0.77$\pm$0.01 & 0.39$\pm$0.23 & 0.50$\pm$0.00 & 0.44$\pm$0.00 & 0.77$\pm$0.01 & 0.00$\pm$0.46 & & & \\
KNeighborsClassifier & 0.60$\pm$0.09 & 0.63$\pm$0.10 & 0.69$\pm$0.07 & 0.58$\pm$0.08 & 0.93$\pm$0.04 & 0.34$\pm$0.20 & & & \\
MLPClassifier & 0.77$\pm$0.27 & 0.47$\pm$0.09 & 0.50$\pm$0.01 & 0.44$\pm$0.13 & 0.77$\pm$0.10 & 0.17$\pm$0.16 & & & \\
GaussianNB & 0.78$\pm$0.22 & 0.39$\pm$0.09 & 0.50$\pm$0.00 & 0.44$\pm$0.10 & 0.78$\pm$0.09 & 0.00$\pm$0.09 & & & \\
AdaBoostClassifier & 0.89$\pm$0.01 & 0.87$\pm$0.03 & 0.79$\pm$0.01 & 0.82$\pm$0.01 & 0.89$\pm$0.01 & 0.85$\pm$0.05 & & & \\
XGB & 0.90$\pm$0.01 & 0.91$\pm$0.01 & 0.80$\pm$0.02 & 0.84$\pm$0.02 & 0.90$\pm$0.01 & 0.93$\pm$0.02 & & & \\
\hline
\multicolumn{7}{c}{2-phase Classification (original sample, N=4,690)} \\ 
\hline
LGBMClassifier & 0.91$\pm$0.01 & 0.93$\pm$0.01 & 0.81$\pm$0.01 & 0.85$\pm$0.01 & 0.90$\pm$0.01 & 0.96$\pm$0.01 & & & \\
LogisticRegression & 0.77$\pm$0.01 & 0.56$\pm$0.07 & 0.50$\pm$0.01 & 0.45$\pm$0.01 & 0.77$\pm$0.01 & 0.36$\pm$0.14 & & & \\
DecisionTreeClassifier & 0.89$\pm$0.01 & 0.87$\pm$0.01 & 0.81$\pm$0.01 & 0.84$\pm$0.01 & 0.91$\pm$0.01 & 0.83$\pm$0.02 & & & \\
RandomForestClassifier & 0.77$\pm$0.01 & 0.64$\pm$0.21 & 0.50$\pm$0.00 & 0.44$\pm$0.01 & 0.77$\pm$0.01 & 0.50$\pm$0.43 & & & \\
KNeighborsClassifier & 0.72$\pm$0.02 & 0.53$\pm$0.01 & 0.51$\pm$0.01 & 0.50$\pm$0.01 & 0.78$\pm$0.01 & 0.28$\pm$0.02 & & & \\
MLPClassifier & 0.77$\pm$0.16 & 0.69$\pm$0.02 & 0.50$\pm$0.05 & 0.44$\pm$0.09 & 0.77$\pm$0.05 & 0.60$\pm$0.06 & & & \\
GaussianNB & 0.77$\pm$0.26 & 0.47$\pm$0.06 & 0.50$\pm$0.01 & 0.44$\pm$0.14 & 0.77$\pm$0.05 & 0.17$\pm$0.12 & & & \\
AdaBoostClassifier & 0.88$\pm$0.01 & 0.85$\pm$0.02 & 0.80$\pm$0.02 & 0.82$\pm$0.02 & 0.90$\pm$0.01 & 0.81$\pm$0.03 & & & \\
XGB & 0.91$\pm$0.01 & 0.92$\pm$0.01 & 0.82$\pm$0.02 & 0.85$\pm$0.02 & 0.90$\pm$0.01 & 0.93$\pm$0.01 & & & \\
\hline
\hline
\end{tabular}
\caption{Numbers of accuracy, precision, recall, F1 score, and individual precision of different classifiers. The uncertainties are derived by bootstrapping.}
\label{tab1}
\end{table*} 

\begin{table*}[t]
\centering
\scriptsize
\begin{tabular}{cccccccccc}
\hline
\hline
data & ACC & P & R & F1 & $P^0$ & $P^1$ & $P^2$ & $P^3$ & $P^4$\\
\hline
\hline
\multicolumn{10}{c}{5-phase Classification (original with rotated sample, N=9,380)} \\ 
\hline
dr+dv+$gri$+H$\alpha$ & 0.85$\pm$0.01 & 0.85$\pm$0.13 & 0.39$\pm$0.01 & 0.40$\pm$0.02 & 0.89$\pm$0.01 & 0.59$\pm$0.03 & 1.00$\pm$0.45 & 0.77$\pm$0.06 & 1.00$\pm$0.49 \\
dr+dv+$gri$ & 0.86$\pm$0.01 & 0.85$\pm$0.10 & 0.40$\pm$0.01 & 0.42$\pm$0.02 & 0.89$\pm$0.01 & 0.61$\pm$0.05 & 1.00$\pm$0.45 & 0.75$\pm$0.07 & 1.00$\pm$0.50 \\
dr+dv+H$\alpha$ & 0.85$\pm$0.01 & 0.68$\pm$0.09 & 0.39$\pm$0.01 & 0.41$\pm$0.02 & 0.89$\pm$0.01 & 0.61$\pm$0.05 & 0.40$\pm$0.32 & 0.52$\pm$0.10 & 1.00$\pm$0.34 \\
$gri$+H$\alpha$ & 0.78$\pm$0.01 & 0.84$\pm$0.14 & 0.22$\pm$0.01 & 0.22$\pm$0.01 & 0.78$\pm$0.01 & 0.40$\pm$0.33 & 1.00$\pm$0.49 & 1.00$\pm$0.30 & 1.00$\pm$0.50 \\
dr+dv & 0.85$\pm$0.01 & 0.71$\pm$0.05 & 0.42$\pm$0.02 & 0.47$\pm$0.02 & 0.89$\pm$0.01 & 0.62$\pm$0.03 & 0.50$\pm$0.14 & 0.55$\pm$0.07 & 1.00$\pm$0.32 \\
dr & 0.84$\pm$0.01 & 0.54$\pm$0.08 & 0.39$\pm$0.02 & 0.43$\pm$0.03 & 0.88$\pm$0.01 & 0.59$\pm$0.03 & 0.30$\pm$0.11 & 0.48$\pm$0.10 & 0.44$\pm$0.38 \\
dv & 0.84$\pm$0.01 & 0.52$\pm$0.04 & 0.39$\pm$0.02 & 0.42$\pm$0.02 & 0.88$\pm$0.01 & 0.60$\pm$0.04 & 0.32$\pm$0.08 & 0.59$\pm$0.09 & 0.19$\pm$0.15 \\
$gri$ & 0.78$\pm$0.01 & 0.89$\pm$0.20 & 0.22$\pm$0.01 & 0.22$\pm$0.01 & 0.78$\pm$0.01 & 0.67$\pm$0.29 & 1.00$\pm$0.30 & 1.00$\pm$0.46 & 1.00$\pm$0.46 \\
$g$ & 0.78$\pm$0.01 & 0.96$\pm$0.18 & 0.22$\pm$0.01 & 0.22$\pm$0.01 & 0.78$\pm$0.01 & 1.00$\pm$0.38 & 1.00$\pm$0.46 & 1.00$\pm$0.49 & 1.00$\pm$0.49 \\
$r$ & 0.78$\pm$0.01 & 0.88$\pm$0.19 & 0.22$\pm$0.01 & 0.22$\pm$0.01 & 0.78$\pm$0.01 & 0.60$\pm$0.39 & 1.00$\pm$0.46 & 1.00$\pm$0.00 & 1.00$\pm$0.49 \\
$i$ & 0.78$\pm$0.01 & 0.81$\pm$0.19 & 0.22$\pm$0.01 & 0.22$\pm$0.02 & 0.78$\pm$0.01 & 0.75$\pm$0.33 & 0.50$\pm$0.49 & 1.00$\pm$0.49 & 1.00$\pm$0.40 \\
H$\alpha$ & 0.77$\pm$0.01 & 0.21$\pm$0.04 & 0.20$\pm$0.00 & 0.18$\pm$0.01 & 0.78$\pm$0.01 & 0.27$\pm$0.14 & 0.00$\pm$0.00 & 0.00$\pm$0.10 & 0.00$\pm$0.00 \\
\hline
\hline
\multicolumn{8}{c}{3-phase Classification (original with rotated sample, N=9,380)} \\ 
\hline
dr+dv+$gri$+H$\alpha$ & 0.89$\pm$0.01 & 0.88$\pm$0.12 & 0.59$\pm$0.01 & 0.62$\pm$0.03 & 0.89$\pm$0.01 & 0.88$\pm$0.03 & 0.88$\pm$0.33 & & \\
dr+dv+$gri$ & 0.89$\pm$0.01 & 0.86$\pm$0.13 & 0.58$\pm$0.01 & 0.60$\pm$0.02 & 0.89$\pm$0.01 & 0.86$\pm$0.02 & 0.83$\pm$0.40 & & \\
dr+dv+H$\alpha$ & 0.88$\pm$0.00 & 0.69$\pm$0.10 & 0.57$\pm$0.01 & 0.59$\pm$0.02 & 0.89$\pm$0.01 & 0.86$\pm$0.03 & 0.33$\pm$0.29 & & \\
$gri$+H$\alpha$ & 0.78$\pm$0.01 & 0.70$\pm$0.15 & 0.35$\pm$0.01 & 0.33$\pm$0.01 & 0.78$\pm$0.01 & 0.73$\pm$0.23 & 0.60$\pm$0.39 & & \\
dr+dv & 0.89$\pm$0.01 & 0.80$\pm$0.05 & 0.60$\pm$0.02 & 0.64$\pm$0.02 & 0.89$\pm$0.00 & 0.90$\pm$0.01 & 0.59$\pm$0.14 & & \\
dr & 0.88$\pm$0.01 & 0.76$\pm$0.05 & 0.58$\pm$0.02 & 0.62$\pm$0.02 & 0.89$\pm$0.01 & 0.89$\pm$0.02 & 0.50$\pm$0.14 & & \\
dv & 0.88$\pm$0.01 & 0.71$\pm$0.03 & 0.57$\pm$0.01 & 0.61$\pm$0.01 & 0.89$\pm$0.01 & 0.89$\pm$0.02 & 0.34$\pm$0.08 & & \\
$gri$ & 0.78$\pm$0.01 & 0.77$\pm$0.11 & 0.35$\pm$0.01 & 0.33$\pm$0.01 & 0.78$\pm$0.01 & 0.53$\pm$0.23 & 1.00$\pm$0.40 & & \\
$g$ & 0.78$\pm$0.01 & 0.85$\pm$0.19 & 0.36$\pm$0.01 & 0.34$\pm$0.02 & 0.78$\pm$0.01 & 0.77$\pm$0.29 & 1.00$\pm$0.46 & & \\
$r$ & 0.78$\pm$0.01 & 0.78$\pm$0.16 & 0.35$\pm$0.01 & 0.34$\pm$0.01 & 0.78$\pm$0.01 & 0.56$\pm$0.21 & 1.00$\pm$0.46 & & \\
$i$ & 0.78$\pm$0.01 & 0.79$\pm$0.18 & 0.35$\pm$0.01 & 0.33$\pm$0.01 & 0.78$\pm$0.01 & 0.60$\pm$0.18 & 1.00$\pm$0.49 & & \\
H$\alpha$ & 0.77$\pm$0.01 & 0.39$\pm$0.12 & 0.34$\pm$0.00 & 0.31$\pm$0.01 & 0.78$\pm$0.01 & 0.39$\pm$0.12 & 0.00$\pm$0.30 & & \\
\hline
\hline
\multicolumn{7}{c}{2-phase Classification (original with rotated sample, N=9,380)} \\ 
\hline
dr+dv+$gri$+H$\alpha$ & 0.90$\pm$0.00 & 0.92$\pm$0.01 & 0.79$\pm$0.01 & 0.84$\pm$0.01 & 0.89$\pm$0.01 & 0.95$\pm$0.01 & & & \\
dr+dv+$gri$ & 0.90$\pm$0.00 & 0.93$\pm$0.00 & 0.79$\pm$0.01 & 0.83$\pm$0.00 & 0.89$\pm$0.01 & 0.96$\pm$0.01 & & & \\
dr+dv+H$\alpha$ & 0.90$\pm$0.01 & 0.92$\pm$0.02 & 0.79$\pm$0.01 & 0.83$\pm$0.01 & 0.89$\pm$0.01 & 0.95$\pm$0.03 & & & \\
$gri$+H$\alpha$ & 0.78$\pm$0.01 & 0.72$\pm$0.05 & 0.54$\pm$0.01 & 0.52$\pm$0.02 & 0.79$\pm$0.01 & 0.65$\pm$0.11 & & & \\
dr+dv & 0.90$\pm$0.01 & 0.93$\pm$0.01 & 0.79$\pm$0.02 & 0.84$\pm$0.02 & 0.89$\pm$0.01 & 0.97$\pm$0.01 & & & \\
dr & 0.90$\pm$0.01 & 0.93$\pm$0.01 & 0.79$\pm$0.02 & 0.83$\pm$0.01 & 0.89$\pm$0.01 & 0.97$\pm$0.01 & & & \\
dv & 0.90$\pm$0.01 & 0.93$\pm$0.01 & 0.79$\pm$0.01 & 0.83$\pm$0.01 & 0.89$\pm$0.00 & 0.97$\pm$0.01 & & & \\
$gri$ & 0.78$\pm$0.01 & 0.72$\pm$0.04 & 0.53$\pm$0.01 & 0.51$\pm$0.01 & 0.79$\pm$0.01 & 0.64$\pm$0.09 & & & \\
$g$ & 0.78$\pm$0.01 & 0.68$\pm$0.07 & 0.52$\pm$0.01 & 0.49$\pm$0.01 & 0.78$\pm$0.01 & 0.58$\pm$0.13 & & & \\
$r$ & 0.78$\pm$0.01 & 0.72$\pm$0.05 & 0.53$\pm$0.01 & 0.51$\pm$0.01 & 0.79$\pm$0.01 & 0.66$\pm$0.09 & & & \\
$i$ & 0.78$\pm$0.01 & 0.66$\pm$0.06 & 0.53$\pm$0.01 & 0.50$\pm$0.01 & 0.78$\pm$0.01 & 0.53$\pm$0.12 & & & \\
H$\alpha$ & 0.77$\pm$0.02 & 0.60$\pm$0.04 & 0.52$\pm$0.01 & 0.50$\pm$0.02 & 0.78$\pm$0.02 & 0.42$\pm$0.07 & & & \\
\hline
\hline
\end{tabular}
\caption{Numbers of accuracy, precision, recall, F1 score, and individual precision for different input data, including projected separation (dr), velocity difference (dv), SDSS $gri$-band images ($gri$), and MaNGA H$\alpha$ velocity map (H$\alpha$), with \texttt{LGBMClassifier}. The uncertainties are derived by bootstrapping.}
\label{tab2}
\end{table*} 


\clearpage
\begin{acknowledgments}
We thank the anonymous referee for providing constructive comments that further improve this manuscript. We also thank S. L. Ellison for helpful suggestions. YYC acknowledge financial support from the Ministry of Science and Technology of Taiwan grant 111-2112-M-005 -018 -MY3 and 109-2112-M-005 -003 -MY3. LL thanks support by the Academia Sinica under the Career Development Award CDA107-M03 and the Ministry of Science \& Technology of Taiwan under the grant MOST 108-2628-M-001-001-MY3. HAP acknowledges support by the Ministry of Science and Technology of Taiwan under grant 110-2112-M-032-020-MY3.  CB gratefully acknowledges support from the Natural Sciences and Engineering Research Council of Canada through their postdoctoral fellowship program.

Funding for the Sloan Digital Sky 
Survey IV has been provided by the 
Alfred P. Sloan Foundation, the U.S. 
Department of Energy Office of 
Science, and the Participating 
Institutions. 

SDSS-IV acknowledges support and 
resources from the Center for High 
Performance Computing at the 
University of Utah. The SDSS 
website is www.sdss.org.

SDSS-IV is managed by the 
Astrophysical Research Consortium 
for the Participating Institutions 
of the SDSS Collaboration including 
the Brazilian Participation Group, 
the Carnegie Institution for Science, 
Carnegie Mellon University, Center for 
Astrophysics | Harvard \& 
Smithsonian, the Chilean Participation 
Group, the French Participation Group, 
Instituto de Astrof\'isica de 
Canarias, The Johns Hopkins 
University, Kavli Institute for the 
Physics and Mathematics of the 
Universe (IPMU) / University of 
Tokyo, the Korean Participation Group, 
Lawrence Berkeley National Laboratory, 
Leibniz Institut f\"ur Astrophysik 
Potsdam (AIP), Max-Planck-Institut 
f\"ur Astronomie (MPIA Heidelberg), 
Max-Planck-Institut f\"ur 
Astrophysik (MPA Garching), 
Max-Planck-Institut f\"ur 
Extraterrestrische Physik (MPE), 
National Astronomical Observatories of 
China, New Mexico State University, 
New York University, University of 
Notre Dame, Observat\'ario 
Nacional / MCTI, The Ohio State 
University, Pennsylvania State 
University, Shanghai 
Astronomical Observatory, United 
Kingdom Participation Group, 
Universidad Nacional Aut\'onoma 
de M\'exico, University of Arizona, 
University of Colorado Boulder, 
University of Oxford, University of 
Portsmouth, University of Utah, 
University of Virginia, University 
of Washington, University of 
Wisconsin, Vanderbilt University, 
and Yale University.
\end{acknowledgments}

\bibliographystyle{aasjournal}

\clearpage
\begin{thebibliography}{}
\expandafter\ifx\csname natexlab\endcsname\relax\def\natexlab#1{#1}\fi
\providecommand{\url}[1]{\href{#1}{#1}}
\providecommand{\dodoi}[1]{doi:~\href{http://doi.org/#1}{\nolinkurl{#1}}}
\providecommand{\doeprint}[1]{\href{http://ascl.net/#1}{\nolinkurl{http://ascl.net/#1}}}
\providecommand{\doarXiv}[1]{\href{https://arxiv.org/abs/#1}{\nolinkurl{https://arxiv.org/abs/#1}}}

\bibitem[{{Abraham} {et~al.}(1994){Abraham}, {Valdes}, {Yee}, \& {van den
  Bergh}}]{1994ApJ...432...75A}
{Abraham}, R.~G., {Valdes}, F., {Yee}, H.~K.~C., \& {van den Bergh}, S. 1994,
  \apj, 432, 75, \dodoi{10.1086/174550}

\bibitem[{{Ackermann} {et~al.}(2018){Ackermann}, {Schawinski}, {Zhang},
  {Weigel}, \& {Turp}}]{2018MNRAS.479..415A}
{Ackermann}, S., {Schawinski}, K., {Zhang}, C., {Weigel}, A.~K., \& {Turp},
  M.~D. 2018, \mnras, 479, 415, \dodoi{10.1093/mnras/sty1398}

\bibitem[{{Albareti} {et~al.}(2017){Albareti}, {Allende Prieto}, {Almeida},
  {Anders}, {Anderson}, {Andrews}, {Arag{\'o}n-Salamanca},
  {Argudo-Fern{\'a}ndez}, {Armengaud}, {Aubourg}, {Avila-Reese}, {Badenes},
  {Bailey}, {Barbuy}, {Barger}, {Barrera-Ballesteros}, {Bartosz}, {Basu},
  {Bates}, {Battaglia}, {Baumgarten}, {Baur}, {Bautista}, {Beers}, {Belfiore},
  {Bershady}, {Bertran de Lis}, {Bird}, {Bizyaev}, {Blanc}, {Blanton},
  {Blomqvist}, {Bolton}, {Borissova}, {Bovy}, {Brandt}, {Brinkmann},
  {Brownstein}, {Bundy}, {Burtin}, {Busca}, {Camacho Chavez}, {Cano D{\'\i}az},
  {Cappellari}, {Carrera}, {Chen}, {Cherinka}, {Cheung}, {Chiappini},
  {Chojnowski}, {Chuang}, {Chung}, {Cirolini}, {Clerc}, {Cohen}, {Comerford},
  {Comparat}, {Correa do Nascimento}, {Cousinou}, {Covey}, {Crane}, {Croft},
  {Cunha}, {Darling}, {Davidson}, {Dawson}, {Da Costa}, {Da Silva Ilha},
  {Deconto Machado}, {Delubac}, {De Lee}, {De la Macorra}, {De la Torre},
  {Diamond-Stanic}, {Donor}, {Downes}, {Drory}, {Du}, {Du Mas des Bourboux},
  {Dwelly}, {Ebelke}, {Eigenbrot}, {Eisenstein}, {Elsworth}, {Emsellem},
  {Eracleous}, {Escoffier}, {Evans}, {Falc{\'o}n-Barroso}, {Fan}, {Favole},
  {Fernandez-Alvar}, {Fernandez-Trincado}, {Feuillet}, {Fleming},
  {Font-Ribera}, {Freischlad}, {Frinchaboy}, {Fu}, {Gao}, {Garcia},
  {Garcia-Dias}, {Garcia-Hern{\'a}ndez}, {Garcia P{\'e}rez}, {Gaulme}, {Ge},
  {Geisler}, {Gillespie}, {Gil Marin}, {Girardi}, {Goddard}, {Gomez Maqueo
  Chew}, {Gonzalez-Perez}, {Grabowski}, {Green}, {Grier}, {Grier}, {Guo},
  {Guy}, {Hagen}, {Hall}, {Harding}, {Harley}, {Hasselquist}, {Hawley},
  {Hayes}, {Hearty}, {Hekker}, {Hernandez Toledo}, {Ho}, {Hogg},
  {Holley-Bockelmann}, {Holtzman}, {Holzer}, {Hu}, {Huber}, {Hutchinson},
  {Hwang}, {Ibarra-Medel}, {Ivans}, {Ivory}, {Jaehnig}, {Jensen}, {Johnson},
  {Jones}, {Jullo}, {Kallinger}, {Kinemuchi}, {Kirkby}, {Klaene}, {Kneib},
  {Kollmeier}, {Lacerna}, {Lane}, {Lang}, {Laurent}, {Law}, {Leauthaud}, {Le
  Goff}, {Li}, {Li}, {Li}, {Li}, {Liang}, {Liang}, {Lima}, {Lin}, {Lin}, {Lin},
  {Liu}, {Long}, {Lucatello}, {MacDonald}, {MacLeod}, {Mackereth}, {Mahadevan},
  {Maia}, {Maiolino}, {Majewski}, {Malanushenko}, {Malanushenko}, {Mallmann},
  {Manchado}, {Maraston}, {Marques-Chaves}, {Martinez Valpuesta}, {Masters},
  {Mathur}, {McGreer}, {Merloni}, {Merrifield}, {M{\'e}sz{\'a}ros}, {Meza},
  {Miglio}, {Minchev}, {Molaverdikhani}, {Montero-Dorta}, {Mosser}, {Muna},
  {Myers}, {Nair}, {Nandra}, {Ness}, {Newman}, {Nichol}, {Nidever},
  {Nitschelm}, {O'Connell}, {Oravetz}, {Oravetz}, {Pace}, {Padilla},
  {Palanque-Delabrouille}, {Pan}, {Parejko}, {Paris}, {Park}, {Peacock},
  {Peirani}, {Pellejero-Ibanez}, {Penny}, {Percival}, {Percival},
  {Perez-Fournon}, {Petitjean}, {Pieri}, {Pinsonneault}, {Pisani}, {Prada},
  {Prakash}, {Price-Jones}, {Raddick}, {Rahman}, {Raichoor}, {Barboza Rembold},
  {Reyna}, {Rich}, {Richstein}, {Ridl}, {Riffel}, {Riffel}, {Rix}, {Robin},
  {Rockosi}, {Rodr{\'\i}guez-Torres}, {Rodrigues}, {Roe}, {Roman Lopes},
  {Rom{\'a}n-Z{\'u}{\~n}iga}, {Ross}, {Rossi}, {Ruan}, {Ruggeri}, {Runnoe},
  {Salazar-Albornoz}, {Salvato}, {Sanchez}, {Sanchez}, {Sanchez-Gallego},
  {Santiago}, {Schiavon}, {Schimoia}, {Schlafly}, {Schlegel}, {Schneider},
  {Sch{\"o}nrich}, {Schultheis}, {Schwope}, {Seo}, {Serenelli}, {Sesar},
  {Shao}, {Shetrone}, {Shull}, {Silva Aguirre}, {Skrutskie}, {Slosar}, {Smith},
  {Smith}, {Sobeck}, {Somers}, {Souto}, {Stark}, {Stassun}, {Steinmetz},
  {Stello}, {Storchi Bergmann}, {Strauss}, {Streblyanska}, {Stringfellow},
  {Suarez}, {Sun}, {Taghizadeh-Popp}, {Tang}, {Tao}, {Tayar}, {Tembe},
  {Thomas}, {Tinker}, {Tojeiro}, {Tremonti}, {Troup}, {Trump}, {Unda-Sanzana},
  {Valenzuela}, {Van den Bosch}, {Vargas-Maga{\~n}a}, {Vazquez}, {Villanova},
  {Vivek}, {Vogt}, {Wake}, {Walterbos}, {Wang}, {Wang}, {Weaver}, {Weijmans},
  {Weinberg}, {Westfall}, {Whelan}, {Wilcots}, {Wild}, {Williams}, {Wilson},
  {Wood-Vasey}, {Wylezalek}, {Xiao}, {Yan}, {Yang}, {Ybarra}, {Yeche}, {Yuan},
  {Zakamska}, {Zamora}, {Zasowski}, {Zhang}, {Zhao}, {Zhao}, {Zheng}, {Zheng},
  {Zhou}, {Zhu}, {Zinn}, \& {Zou}}]{2017ApJS..233...25A}
{Albareti}, F.~D., {Allende Prieto}, C., {Almeida}, A., {et~al.} 2017, \apjs,
  233, 25, \dodoi{10.3847/1538-4365/aa8992}

\bibitem[{{Banerji} {et~al.}(2010){Banerji}, {Lahav}, {Lintott}, {Abdalla},
  {Schawinski}, {Bamford}, {Andreescu}, {Murray}, {Raddick}, {Slosar},
  {Szalay}, {Thomas}, \& {Vandenberg}}]{2010MNRAS.406..342B}
{Banerji}, M., {Lahav}, O., {Lintott}, C.~J., {et~al.} 2010, \mnras, 406, 342,
  \dodoi{10.1111/j.1365-2966.2010.16713.x}

\bibitem[{{Barchi} {et~al.}(2020){Barchi}, {de Carvalho}, {Rosa}, {Sautter},
  {Soares-Santos}, {Marques}, {Clua}, {Gon{\c{c}}alves}, {de S{\'a}-Freitas},
  \& {Moura}}]{2020A&C....3000334B}
{Barchi}, P.~H., {de Carvalho}, R.~R., {Rosa}, R.~R., {et~al.} 2020, Astronomy
  and Computing, 30, 100334, \dodoi{10.1016/j.ascom.2019.100334}

\bibitem[{{Bell} {et~al.}(2012){Bell}, {van der Wel}, {Papovich}, {Kocevski},
  {Lotz}, {McIntosh}, {Kartaltepe}, {Faber}, {Ferguson}, {Koekemoer}, {Grogin},
  {Wuyts}, {Cheung}, {Conselice}, {Dekel}, {Dunlop}, {Giavalisco},
  {Herrington}, {Koo}, {McGrath}, {de Mello}, {Rix}, {Robaina}, \&
  {Williams}}]{2012ApJ...753..167B}
{Bell}, E.~F., {van der Wel}, A., {Papovich}, C., {et~al.} 2012, \apj, 753,
  167, \dodoi{10.1088/0004-637X/753/2/167}

\bibitem[{{Bickley} {et~al.}(2022){Bickley}, {Ellison}, {Patton}, {Bottrell},
  {Gwyn}, \& {Hudson}}]{2022MNRAS.514.3294B}
{Bickley}, R.~W., {Ellison}, S.~L., {Patton}, D.~R., {et~al.} 2022, \mnras,
  514, 3294, \dodoi{10.1093/mnras/stac1500}

\bibitem[{{Bickley} {et~al.}(2021){Bickley}, {Bottrell}, {Hani}, {Ellison},
  {Teimoorinia}, {Yi}, {Wilkinson}, {Gwyn}, \& {Hudson}}]{2021MNRAS.504..372B}
{Bickley}, R.~W., {Bottrell}, C., {Hani}, M.~H., {et~al.} 2021, \mnras, 504,
  372, \dodoi{10.1093/mnras/stab806}

\bibitem[{{Blanton} {et~al.}(2017){Blanton}, {Bershady}, {Abolfathi},
  {Albareti}, {Allende Prieto}, {Almeida}, {Alonso-Garc{\'\i}a}, {Anders},
  {Anderson}, {Andrews}, {Aquino-Ort{\'\i}z}, {Arag{\'o}n-Salamanca},
  {Argudo-Fern{\'a}ndez}, {Armengaud}, {Aubourg}, {Avila-Reese}, {Badenes},
  {Bailey}, {Barger}, {Barrera-Ballesteros}, {Bartosz}, {Bates}, {Baumgarten},
  {Bautista}, {Beaton}, {Beers}, {Belfiore}, {Bender}, {Berlind}, {Bernardi},
  {Beutler}, {Bird}, {Bizyaev}, {Blanc}, {Blomqvist}, {Bolton}, {Boquien},
  {Borissova}, {van den Bosch}, {Bovy}, {Brandt}, {Brinkmann}, {Brownstein},
  {Bundy}, {Burgasser}, {Burtin}, {Busca}, {Cappellari}, {Delgado Carigi},
  {Carlberg}, {Carnero Rosell}, {Carrera}, {Chanover}, {Cherinka}, {Cheung},
  {G{\'o}mez Maqueo Chew}, {Chiappini}, {Choi}, {Chojnowski}, {Chuang},
  {Chung}, {Cirolini}, {Clerc}, {Cohen}, {Comparat}, {da Costa}, {Cousinou},
  {Covey}, {Crane}, {Croft}, {Cruz-Gonzalez}, {Garrido Cuadra}, {Cunha},
  {Damke}, {Darling}, {Davies}, {Dawson}, {de la Macorra}, {Dell'Agli}, {De
  Lee}, {Delubac}, {Di Mille}, {Diamond-Stanic}, {Cano-D{\'\i}az}, {Donor},
  {Downes}, {Drory}, {du Mas des Bourboux}, {Duckworth}, {Dwelly}, {Dyer},
  {Ebelke}, {Eigenbrot}, {Eisenstein}, {Emsellem}, {Eracleous}, {Escoffier},
  {Evans}, {Fan}, {Fern{\'a}ndez-Alvar}, {Fernandez-Trincado}, {Feuillet},
  {Finoguenov}, {Fleming}, {Font-Ribera}, {Fredrickson}, {Freischlad},
  {Frinchaboy}, {Fuentes}, {Galbany}, {Garcia-Dias},
  {Garc{\'\i}a-Hern{\'a}ndez}, {Gaulme}, {Geisler}, {Gelfand},
  {Gil-Mar{\'\i}n}, {Gillespie}, {Goddard}, {Gonzalez-Perez}, {Grabowski},
  {Green}, {Grier}, {Gunn}, {Guo}, {Guy}, {Hagen}, {Hahn}, {Hall}, {Harding},
  {Hasselquist}, {Hawley}, {Hearty}, {Gonzalez Hern{\'a}ndez}, {Ho}, {Hogg},
  {Holley-Bockelmann}, {Holtzman}, {Holzer}, {Huehnerhoff}, {Hutchinson},
  {Hwang}, {Ibarra-Medel}, {da Silva Ilha}, {Ivans}, {Ivory}, {Jackson},
  {Jensen}, {Johnson}, {Jones}, {J{\"o}nsson}, {Jullo}, {Kamble}, {Kinemuchi},
  {Kirkby}, {Kitaura}, {Klaene}, {Knapp}, {Kneib}, {Kollmeier}, {Lacerna},
  {Lane}, {Lang}, {Law}, {Lazarz}, {Lee}, {Le Goff}, {Liang}, {Li}, {Li},
  {Lian}, {Lima}, {Lin}, {Lin}, {Bertran de Lis}, {Liu}, {de Icaza Lizaola},
  {Long}, {Lucatello}, {Lundgren}, {MacDonald}, {Deconto Machado}, {MacLeod},
  {Mahadevan}, {Geimba Maia}, {Maiolino}, {Majewski}, {Malanushenko},
  {Malanushenko}, {Manchado}, {Mao}, {Maraston}, {Marques-Chaves}, {Masseron},
  {Masters}, {McBride}, {McDermid}, {McGrath}, {McGreer}, {Medina Pe{\~n}a},
  {Melendez}, {Merloni}, {Merrifield}, {Meszaros}, {Meza}, {Minchev},
  {Minniti}, {Miyaji}, {More}, {Mulchaey}, {M{\"u}ller-S{\'a}nchez}, {Muna},
  {Munoz}, {Myers}, {Nair}, {Nandra}, {Correa do Nascimento}, {Negrete},
  {Ness}, {Newman}, {Nichol}, {Nidever}, {Nitschelm}, {Ntelis}, {O'Connell},
  {Oelkers}, {Oravetz}, {Oravetz}, {Pace}, {Padilla}, {Palanque-Delabrouille},
  {Alonso Palicio}, {Pan}, {Parejko}, {Parikh}, {P{\^a}ris}, {Park}, {Patten},
  {Peirani}, {Pellejero-Ibanez}, {Penny}, {Percival}, {Perez-Fournon},
  {Petitjean}, {Pieri}, {Pinsonneault}, {Pisani}, {Poleski}, {Prada},
  {Prakash}, {Queiroz}, {Raddick}, {Raichoor}, {Barboza Rembold}, {Richstein},
  {Riffel}, {Riffel}, {Rix}, {Robin}, {Rockosi}, {Rodr{\'\i}guez-Torres},
  {Roman-Lopes}, {Rom{\'a}n-Z{\'u}{\~n}iga}, {Rosado}, {Ross}, {Rossi}, {Ruan},
  {Ruggeri}, {Rykoff}, {Salazar-Albornoz}, {Salvato}, {S{\'a}nchez}, {Aguado},
  {S{\'a}nchez-Gallego}, {Santana}, {Santiago}, {Sayres}, {Schiavon}, {da Silva
  Schimoia}, {Schlafly}, {Schlegel}, {Schneider}, {Schultheis}, {Schuster},
  {Schwope}, {Seo}, {Shao}, {Shen}, {Shetrone}, {Shull}, {Simon}, {Skinner},
  {Skrutskie}, {Slosar}, {Smith}, {Sobeck}, {Sobreira}, {Somers}, {Souto},
  {Stark}, {Stassun}, {Stauffer}, {Steinmetz}, {Storchi-Bergmann},
  {Streblyanska}, {Stringfellow}, {Su{\'a}rez}, {Sun}, {Suzuki}, {Szigeti},
  {Taghizadeh-Popp}, {Tang}, {Tao}, {Tayar}, {Tembe}, {Teske}, {Thakar},
  {Thomas}, {Thompson}, {Tinker}, {Tissera}, {Tojeiro}, {Hernandez Toledo}, {de
  la Torre}, {Tremonti}, {Troup}, {Valenzuela}, {Martinez Valpuesta},
  {Vargas-Gonz{\'a}lez}, {Vargas-Maga{\~n}a}, {Vazquez}, {Villanova}, {Vivek},
  {Vogt}, {Wake}, {Walterbos}, {Wang}, {Weaver}, {Weijmans}, {Weinberg},
  {Westfall}, {Whelan}, {Wild}, {Wilson}, {Wood-Vasey}, {Wylezalek}, {Xiao},
  {Yan}, {Yang}, {Ybarra}, {Y{\`e}che}, {Zakamska}, {Zamora}, {Zarrouk},
  {Zasowski}, {Zhang}, {Zhao}, {Zheng}, {Zheng}, {Zhou}, {Zhou}, {Zhu},
  {Zoccali}, \& {Zou}}]{2017AJ....154...28B}
{Blanton}, M.~R., {Bershady}, M.~A., {Abolfathi}, B., {et~al.} 2017, \aj, 154,
  28, \dodoi{10.3847/1538-3881/aa7567}

\bibitem[{{Bonjean} {et~al.}(2019){Bonjean}, {Aghanim}, {Salom{\'e}}, {Beelen},
  {Douspis}, \& {Soubri{\'e}}}]{2019A&A...622A.137B}
{Bonjean}, V., {Aghanim}, N., {Salom{\'e}}, P., {et~al.} 2019, \aap, 622, A137,
  \dodoi{10.1051/0004-6361/201833972}

\bibitem[{{Bottrell} {et~al.}(2022){Bottrell}, {Hani}, {Teimoorinia}, {Patton},
  \& {Ellison}}]{2022MNRAS.511..100B}
{Bottrell}, C., {Hani}, M.~H., {Teimoorinia}, H., {Patton}, D.~R., \&
  {Ellison}, S.~L. 2022, \mnras, 511, 100, \dodoi{10.1093/mnras/stab3717}

\bibitem[{{Bottrell} {et~al.}(2019){Bottrell}, {Hani}, {Teimoorinia},
  {Ellison}, {Moreno}, {Torrey}, {Hayward}, {Thorp}, {Simard}, \&
  {Hernquist}}]{2019MNRAS.490.5390B}
{Bottrell}, C., {Hani}, M.~H., {Teimoorinia}, H., {et~al.} 2019, \mnras, 490,
  5390, \dodoi{10.1093/mnras/stz2934}

\bibitem[{{Bryant} {et~al.}(2015){Bryant}, {Owers}, {Robotham}, {Croom},
  {Driver}, {Drinkwater}, {Lorente}, {Cortese}, {Scott}, {Colless}, {Schaefer},
  {Taylor}, {Konstantopoulos}, {Allen}, {Baldry}, {Barnes}, {Bauer},
  {Bland-Hawthorn}, {Bloom}, {Brooks}, {Brough}, {Cecil}, {Couch}, {Croton},
  {Davies}, {Ellis}, {Fogarty}, {Foster}, {Glazebrook}, {Goodwin}, {Green},
  {Gunawardhana}, {Hampton}, {Ho}, {Hopkins}, {Kewley}, {Lawrence},
  {Leon-Saval}, {Leslie}, {McElroy}, {Lewis}, {Liske}, {L{\'o}pez-S{\'a}nchez},
  {Mahajan}, {Medling}, {Metcalfe}, {Meyer}, {Mould}, {Obreschkow}, {O'Toole},
  {Pracy}, {Richards}, {Shanks}, {Sharp}, {Sweet}, {Thomas}, {Tonini}, \&
  {Walcher}}]{2015MNRAS.447.2857B}
{Bryant}, J.~J., {Owers}, M.~S., {Robotham}, A.~S.~G., {et~al.} 2015, \mnras,
  447, 2857, \dodoi{10.1093/mnras/stu2635}

\bibitem[{{Bryant} {et~al.}(2016){Bryant}, {Bland-Hawthorn}, {Lawrence},
  {Croom}, {Brown}, {Venkatesan}, {Gillingham}, {Zhelem}, {Content},
  {Saunders}, {Staszak}, {van de Sande}, {Couch}, {Leon-Saval}, {Tims},
  {McDermid}, \& {Schaefer}}]{2016SPIE.9908E..1FB}
{Bryant}, J.~J., {Bland-Hawthorn}, J., {Lawrence}, J., {et~al.} 2016, in
  Society of Photo-Optical Instrumentation Engineers (SPIE) Conference Series,
  Vol. 9908, Ground-based and Airborne Instrumentation for Astronomy VI, ed.
  C.~J. {Evans}, L.~{Simard}, \& H.~{Takami}, 99081F,
  \dodoi{10.1117/12.2230740}

\bibitem[{{Bundy} {et~al.}(2015){Bundy}, {Bershady}, {Law}, {Yan}, {Drory},
  {MacDonald}, {Wake}, {Cherinka}, {S{\'a}nchez-Gallego}, {Weijmans}, {Thomas},
  {Tremonti}, {Masters}, {Coccato}, {Diamond-Stanic}, {Arag{\'o}n-Salamanca},
  {Avila-Reese}, {Badenes}, {Falc{\'o}n-Barroso}, {Belfiore}, {Bizyaev},
  {Blanc}, {Bland-Hawthorn}, {Blanton}, {Brownstein}, {Byler}, {Cappellari},
  {Conroy}, {Dutton}, {Emsellem}, {Etherington}, {Frinchaboy}, {Fu}, {Gunn},
  {Harding}, {Johnston}, {Kauffmann}, {Kinemuchi}, {Klaene}, {Knapen},
  {Leauthaud}, {Li}, {Lin}, {Maiolino}, {Malanushenko}, {Malanushenko}, {Mao},
  {Maraston}, {McDermid}, {Merrifield}, {Nichol}, {Oravetz}, {Pan}, {Parejko},
  {Sanchez}, {Schlegel}, {Simmons}, {Steele}, {Steinmetz}, {Thanjavur},
  {Thompson}, {Tinker}, {van den Bosch}, {Westfall}, {Wilkinson}, {Wright},
  {Xiao}, \& {Zhang}}]{2015ApJ...798....7B}
{Bundy}, K., {Bershady}, M.~A., {Law}, D.~R., {et~al.} 2015, \apj, 798, 7,
  \dodoi{10.1088/0004-637X/798/1/7}

\bibitem[{{Cappellari} {et~al.}(2011){Cappellari}, {Emsellem}, {Krajnovi{\'c}},
  {McDermid}, {Scott}, {Verdoes Kleijn}, {Young}, {Alatalo}, {Bacon}, {Blitz},
  {Bois}, {Bournaud}, {Bureau}, {Davies}, {Davis}, {de Zeeuw}, {Duc},
  {Khochfar}, {Kuntschner}, {Lablanche}, {Morganti}, {Naab}, {Oosterloo},
  {Sarzi}, {Serra}, \& {Weijmans}}]{2011MNRAS.413..813C}
{Cappellari}, M., {Emsellem}, E., {Krajnovi{\'c}}, D., {et~al.} 2011, \mnras,
  413, 813, \dodoi{10.1111/j.1365-2966.2010.18174.x}

\bibitem[{{Chabrier}(2003)}]{2003PASP..115..763C}
{Chabrier}, G. 2003, \pasp, 115, 763, \dodoi{10.1086/376392}

\bibitem[{{Chang} {et~al.}(2021){Chang}, {Hsieh}, {Wang}, {Lin}, {Lim}, {Toba},
  {Zhong}, \& {Chang}}]{2021ApJ...920...68C}
{Chang}, Y.-Y., {Hsieh}, B.-C., {Wang}, W.-H., {et~al.} 2021, \apj, 920, 68,
  \dodoi{10.3847/1538-4357/ac167c}

\bibitem[{Chen \& Guestrin(2016)}]{Chen:2016:XST:2939672.2939785}
Chen, T., \& Guestrin, C. 2016, in Proceedings of the 22nd ACM SIGKDD
  International Conference on Knowledge Discovery and Data Mining, KDD '16 (New
  York, NY, USA: ACM), 785--794, \dodoi{10.1145/2939672.2939785}

\bibitem[{{{\'C}iprijanovi{\'c}} {et~al.}(2021){{\'C}iprijanovi{\'c}},
  {Kafkes}, {Downey}, {Jenkins}, {Perdue}, {Madireddy}, {Johnston}, {Snyder},
  \& {Nord}}]{2021MNRAS.506..677C}
{{\'C}iprijanovi{\'c}}, A., {Kafkes}, D., {Downey}, K., {et~al.} 2021, \mnras,
  506, 677, \dodoi{10.1093/mnras/stab1677}

\bibitem[{{Conselice}(2003)}]{2003ApJS..147....1C}
{Conselice}, C.~J. 2003, \apjs, 147, 1, \dodoi{10.1086/375001}

\bibitem[{{Conselice}(2014)}]{2014ARA&A..52..291C}
---. 2014, \araa, 52, 291, \dodoi{10.1146/annurev-astro-081913-040037}

\bibitem[{{Davidzon} {et~al.}(2019){Davidzon}, {Laigle}, {Capak}, {Ilbert},
  {Masters}, {Hemmati}, {Apostolakos}, {Coupon}, {de la Torre}, {Devriendt},
  {Dubois}, {Kashino}, {Paltani}, \& {Pichon}}]{2019MNRAS.489.4817D}
{Davidzon}, I., {Laigle}, C., {Capak}, P.~L., {et~al.} 2019, \mnras, 489, 4817,
  \dodoi{10.1093/mnras/stz2486}

\bibitem[{{D'Isanto} \& {Polsterer}(2018)}]{2018A&A...609A.111D}
{D'Isanto}, A., \& {Polsterer}, K.~L. 2018, \aap, 609, A111,
  \dodoi{10.1051/0004-6361/201731326}

\bibitem[{{Dom{\'\i}nguez S{\'a}nchez} {et~al.}(2018){Dom{\'\i}nguez
  S{\'a}nchez}, {Huertas-Company}, {Bernardi}, {Tuccillo}, \&
  {Fischer}}]{2018MNRAS.476.3661D}
{Dom{\'\i}nguez S{\'a}nchez}, H., {Huertas-Company}, M., {Bernardi}, M.,
  {Tuccillo}, D., \& {Fischer}, J.~L. 2018, \mnras, 476, 3661,
  \dodoi{10.1093/mnras/sty338}

\bibitem[{{Donnari} {et~al.}(2019){Donnari}, {Pillepich}, {Nelson},
  {Vogelsberger}, {Genel}, {Weinberger}, {Marinacci}, {Springel}, \&
  {Hernquist}}]{2019MNRAS.485.4817D}
{Donnari}, M., {Pillepich}, A., {Nelson}, D., {et~al.} 2019, \mnras, 485, 4817,
  \dodoi{10.1093/mnras/stz712}

\bibitem[{{Drory} {et~al.}(2015){Drory}, {MacDonald}, {Bershady}, {Bundy},
  {Gunn}, {Law}, {Smith}, {Stoll}, {Tremonti}, {Wake}, {Yan}, {Weijmans},
  {Byler}, {Cherinka}, {Cope}, {Eigenbrot}, {Harding}, {Holder}, {Huehnerhoff},
  {Jaehnig}, {Jansen}, {Klaene}, {Paat}, {Percival}, \&
  {Sayres}}]{2015AJ....149...77D}
{Drory}, N., {MacDonald}, N., {Bershady}, M.~A., {et~al.} 2015, \aj, 149, 77,
  \dodoi{10.1088/0004-6256/149/2/77}

\bibitem[{{Ellison} {et~al.}(2013){Ellison}, {Mendel}, {Patton}, \&
  {Scudder}}]{2013MNRAS.435.3627E}
{Ellison}, S.~L., {Mendel}, J.~T., {Patton}, D.~R., \& {Scudder}, J.~M. 2013,
  \mnras, 435, 3627, \dodoi{10.1093/mnras/stt1562}

\bibitem[{{Ellison} {et~al.}(2008){Ellison}, {Patton}, {Simard}, \&
  {McConnachie}}]{2008AJ....135.1877E}
{Ellison}, S.~L., {Patton}, D.~R., {Simard}, L., \& {McConnachie}, A.~W. 2008,
  \aj, 135, 1877, \dodoi{10.1088/0004-6256/135/5/1877}

\bibitem[{{Ellison} {et~al.}(2018){Ellison}, {S{\'a}nchez}, {Ibarra-Medel},
  {Antonio}, {Mendel}, \& {Barrera-Ballesteros}}]{2018MNRAS.474.2039E}
{Ellison}, S.~L., {S{\'a}nchez}, S.~F., {Ibarra-Medel}, H., {et~al.} 2018,
  \mnras, 474, 2039, \dodoi{10.1093/mnras/stx2882}

\bibitem[{{Faber} {et~al.}(2007){Faber}, {Willmer}, {Wolf}, {Koo}, {Weiner},
  {Newman}, {Im}, {Coil}, {Conroy}, {Cooper}, {Davis}, {Finkbeiner}, {Gerke},
  {Gebhardt}, {Groth}, {Guhathakurta}, {Harker}, {Kaiser}, {Kassin},
  {Kleinheinrich}, {Konidaris}, {Kron}, {Lin}, {Luppino}, {Madgwick},
  {Meisenheimer}, {Noeske}, {Phillips}, {Sarajedini}, {Schiavon}, {Simard},
  {Szalay}, {Vogt}, \& {Yan}}]{2007ApJ...665..265F}
{Faber}, S.~M., {Willmer}, C.~N.~A., {Wolf}, C., {et~al.} 2007, \apj, 665, 265,
  \dodoi{10.1086/519294}

\bibitem[{{Ferreira} {et~al.}(2020){Ferreira}, {Conselice}, {Duncan}, {Cheng},
  {Griffiths}, \& {Whitney}}]{2020ApJ...895..115F}
{Ferreira}, L., {Conselice}, C.~J., {Duncan}, K., {et~al.} 2020, \apj, 895,
  115, \dodoi{10.3847/1538-4357/ab8f9b}

\bibitem[{{Ferreira} {et~al.}(2022){Ferreira}, {Conselice}, {Kuchner}, \&
  {Tohill}}]{2022ApJ...931...34F}
{Ferreira}, L., {Conselice}, C.~J., {Kuchner}, U., \& {Tohill}, C.-B. 2022,
  \apj, 931, 34, \dodoi{10.3847/1538-4357/ac66ea}

\bibitem[{{Gonz{\'a}lez Delgado} {et~al.}(2014){Gonz{\'a}lez Delgado},
  {P{\'e}rez}, {Cid Fernandes}, {Garc{\'\i}a-Benito}, {de Amorim},
  {S{\'a}nchez}, {Husemann}, {Cortijo-Ferrero}, {L{\'o}pez Fern{\'a}ndez},
  {S{\'a}nchez-Bl{\'a}zquez}, {Bekeraite}, {Walcher}, {Falc{\'o}n-Barroso},
  {Gallazzi}, {van de Ven}, {Alves}, {Bland-Hawthorn}, {Kennicutt}, {Kupko},
  {Lyubenova}, {Mast}, {Moll{\'a}}, {Marino}, {Quirrenbach}, {V{\'\i}lchez}, \&
  {Wisotzki}}]{2014A&A...562A..47G}
{Gonz{\'a}lez Delgado}, R.~M., {P{\'e}rez}, E., {Cid Fernandes}, R., {et~al.}
  2014, \aap, 562, A47, \dodoi{10.1051/0004-6361/201322011}

\bibitem[{{Gunn} {et~al.}(2006){Gunn}, {Siegmund}, {Mannery}, {Owen}, {Hull},
  {Leger}, {Carey}, {Knapp}, {York}, {Boroski}, {Kent}, {Lupton}, {Rockosi},
  {Evans}, {Waddell}, {Anderson}, {Annis}, {Barentine}, {Bartoszek}, {Bastian},
  {Bracker}, {Brewington}, {Briegel}, {Brinkmann}, {Brown}, {Carr},
  {Czarapata}, {Drennan}, {Dombeck}, {Federwitz}, {Gillespie}, {Gonzales},
  {Hansen}, {Harvanek}, {Hayes}, {Jordan}, {Kinney}, {Klaene}, {Kleinman},
  {Kron}, {Kresinski}, {Lee}, {Limmongkol}, {Lindenmeyer}, {Long}, {Loomis},
  {McGehee}, {Mantsch}, {Neilsen}, {Neswold}, {Newman}, {Nitta}, {Peoples},
  {Pier}, {Prieto}, {Prosapio}, {Rivetta}, {Schneider}, {Snedden}, \&
  {Wang}}]{2006AJ....131.2332G}
{Gunn}, J.~E., {Siegmund}, W.~A., {Mannery}, E.~J., {et~al.} 2006, \aj, 131,
  2332, \dodoi{10.1086/500975}

\bibitem[{{Hemmati} {et~al.}(2019){Hemmati}, {Capak}, {Pourrahmani}, {Nayyeri},
  {Stern}, {Mobasher}, {Darvish}, {Davidzon}, {Ilbert}, {Masters}, \&
  {Shahidi}}]{2019ApJ...881L..14H}
{Hemmati}, S., {Capak}, P., {Pourrahmani}, M., {et~al.} 2019, \apjl, 881, L14,
  \dodoi{10.3847/2041-8213/ab3418}

\bibitem[{{Hopkins} {et~al.}(2006){Hopkins}, {Hernquist}, {Cox}, {Di Matteo},
  {Robertson}, \& {Springel}}]{2006ApJS..163....1H}
{Hopkins}, P.~F., {Hernquist}, L., {Cox}, T.~J., {et~al.} 2006, \apjs, 163, 1,
  \dodoi{10.1086/499298}

\bibitem[{{Hopkins} {et~al.}(2008){Hopkins}, {Hernquist}, {Cox}, \& {Kere{\v
  s}}}]{2008ApJS..175..356H}
{Hopkins}, P.~F., {Hernquist}, L., {Cox}, T.~J., \& {Kere{\v s}}, D. 2008,
  \apjs, 175, 356, \dodoi{10.1086/524362}

\bibitem[{{Hsieh} {et~al.}(2017){Hsieh}, {Lin}, {Lin}, {Pan}, {Hsu},
  {S{\'a}nchez}, {Cano-D{\'\i}az}, {Zhang}, {Yan}, {Barrera-Ballesteros},
  {Boquien}, {Riffel}, {Brownstein}, {Cruz-Gonz{\'a}lez}, {Hagen}, {Ibarra},
  {Pan}, {Bizyaev}, {Oravetz}, \& {Simmons}}]{2017ApJ...851L..24H}
{Hsieh}, B.~C., {Lin}, L., {Lin}, J.~H., {et~al.} 2017, \apjl, 851, L24,
  \dodoi{10.3847/2041-8213/aa9d80}

\bibitem[{{Huertas-Company} {et~al.}(2015){Huertas-Company}, {Gravet},
  {Cabrera-Vives}, {P{\'e}rez-Gonz{\'a}lez}, {Kartaltepe}, {Barro}, {Bernardi},
  {Mei}, {Shankar}, {Dimauro}, {Bell}, {Kocevski}, {Koo}, {Faber}, \&
  {Mcintosh}}]{2015ApJS..221....8H}
{Huertas-Company}, M., {Gravet}, R., {Cabrera-Vives}, G., {et~al.} 2015, \apjs,
  221, 8, \dodoi{10.1088/0067-0049/221/1/8}

\bibitem[{{Jian} {et~al.}(2018){Jian}, {Lin}, {Oguri}, {Nishizawa}, {Takada},
  {More}, {Koyama}, {Tanaka}, \& {Komiyama}}]{2018PASJ...70S..23J}
{Jian}, H.-Y., {Lin}, L., {Oguri}, M., {et~al.} 2018, \pasj, 70, S23,
  \dodoi{10.1093/pasj/psx096}

\bibitem[{{Kauffmann} \& {Haehnelt}(2000)}]{2000MNRAS.311..576K}
{Kauffmann}, G., \& {Haehnelt}, M. 2000, \mnras, 311, 576,
  \dodoi{10.1046/j.1365-8711.2000.03077.x}

\bibitem[{{Knapen} {et~al.}(2015){Knapen}, {Cisternas}, \&
  {Querejeta}}]{2015MNRAS.454.1742K}
{Knapen}, J.~H., {Cisternas}, M., \& {Querejeta}, M. 2015, \mnras, 454, 1742,
  \dodoi{10.1093/mnras/stv2135}

\bibitem[{{Krakowski} {et~al.}(2016){Krakowski}, {Ma{\l}ek}, {Bilicki},
  {Pollo}, {Kurcz}, \& {Krupa}}]{2016A&A...596A..39K}
{Krakowski}, T., {Ma{\l}ek}, K., {Bilicki}, M., {et~al.} 2016, \aap, 596, A39,
  \dodoi{10.1051/0004-6361/201629165}

\bibitem[{{Law} {et~al.}(2015){Law}, {Yan}, {Bershady}, {Bundy}, {Cherinka},
  {Drory}, {MacDonald}, {S{\'a}nchez-Gallego}, {Wake}, {Weijmans}, {Blanton},
  {Klaene}, {Moran}, {Sanchez}, \& {Zhang}}]{2015AJ....150...19L}
{Law}, D.~R., {Yan}, R., {Bershady}, M.~A., {et~al.} 2015, \aj, 150, 19,
  \dodoi{10.1088/0004-6256/150/1/19}

\bibitem[{{Lee} {et~al.}(2015){Lee}, {Sanders}, {Casey}, {Toft}, {Scoville},
  {Hung}, {Le Floc'h}, {Ilbert}, {Zahid}, {Aussel}, {Capak}, {Kartaltepe},
  {Kewley}, {Li}, {Schawinski}, {Sheth}, \& {Xiao}}]{2015ApJ...801...80L}
{Lee}, N., {Sanders}, D.~B., {Casey}, C.~M., {et~al.} 2015, \apj, 801, 80,
  \dodoi{10.1088/0004-637X/801/2/80}

\bibitem[{{Lin} {et~al.}(2007){Lin}, {Koo}, {Weiner}, {Chiueh}, {Coil}, {Lotz},
  {Conselice}, {Willner}, {Smith}, {Guhathakurta}, {Huang}, {Le Floc'h},
  {Noeske}, {Willmer}, {Cooper}, \& {Phillips}}]{2007ApJ...660L..51L}
{Lin}, L., {Koo}, D.~C., {Weiner}, B.~J., {et~al.} 2007, \apjl, 660, L51,
  \dodoi{10.1086/517919}

\bibitem[{{Lin} {et~al.}(2014){Lin}, {Jian}, {Foucaud}, {Norberg}, {Bower},
  {Cole}, {Arnalte-Mur}, {Chen}, {Coupon}, {Hsieh}, {Heinis}, {Phleps}, {Chen},
  {Lee}, {Burgett}, {Chambers}, {Denneau}, {Draper}, {Flewelling}, {Hodapp},
  {Huber}, {Kaiser}, {Kudritzki}, {Magnier}, {Metcalfe}, {Price}, {Tonry},
  {Wainscoat}, \& {Waters}}]{2014ApJ...782...33L}
{Lin}, L., {Jian}, H.-Y., {Foucaud}, S., {et~al.} 2014, \apj, 782, 33,
  \dodoi{10.1088/0004-637X/782/1/33}

\bibitem[{{Lin} {et~al.}(2019){Lin}, {Pan}, {Ellison}, {Belfiore}, {Shi},
  {S{\'a}nchez}, {Hsieh}, {Rowlands}, {Ramya}, {Thorp}, {Li}, \&
  {Maiolino}}]{2019ApJ...884L..33L}
{Lin}, L., {Pan}, H.-A., {Ellison}, S.~L., {et~al.} 2019, \apjl, 884, L33,
  \dodoi{10.3847/2041-8213/ab4815}

\bibitem[{{Lintott} {et~al.}(2008){Lintott}, {Schawinski}, {Slosar}, {Land},
  {Bamford}, {Thomas}, {Raddick}, {Nichol}, {Szalay}, {Andreescu}, {Murray}, \&
  {Vandenberg}}]{2008MNRAS.389.1179L}
{Lintott}, C.~J., {Schawinski}, K., {Slosar}, A., {et~al.} 2008, \mnras, 389,
  1179, \dodoi{10.1111/j.1365-2966.2008.13689.x}

\bibitem[{{Lotz} {et~al.}(2011){Lotz}, {Jonsson}, {Cox}, {Croton}, {Primack},
  {Somerville}, \& {Stewart}}]{2011ApJ...742..103L}
{Lotz}, J.~M., {Jonsson}, P., {Cox}, T.~J., {et~al.} 2011, \apj, 742, 103,
  \dodoi{10.1088/0004-637X/742/2/103}

\bibitem[{{Lotz} {et~al.}(2010){Lotz}, {Jonsson}, {Cox}, \&
  {Primack}}]{2010MNRAS.404..590L}
{Lotz}, J.~M., {Jonsson}, P., {Cox}, T.~J., \& {Primack}, J.~R. 2010, \mnras,
  404, 590, \dodoi{10.1111/j.1365-2966.2010.16269.x}

\bibitem[{{Lotz} {et~al.}(2004){Lotz}, {Primack}, \&
  {Madau}}]{2004AJ....128..163L}
{Lotz}, J.~M., {Primack}, J., \& {Madau}, P. 2004, \aj, 128, 163,
  \dodoi{10.1086/421849}

\bibitem[{{Masters} {et~al.}(2015){Masters}, {Capak}, {Stern}, {Ilbert},
  {Salvato}, {Schmidt}, {Longo}, {Rhodes}, {Paltani}, {Mobasher}, {Hoekstra},
  {Hildebrandt}, {Coupon}, {Steinhardt}, {Speagle}, {Faisst}, {Kalinich},
  {Brodwin}, {Brescia}, \& {Cavuoti}}]{2015ApJ...813...53M}
{Masters}, D., {Capak}, P., {Stern}, D., {et~al.} 2015, \apj, 813, 53,
  \dodoi{10.1088/0004-637X/813/1/53}

\bibitem[{{McElroy} {et~al.}(2022){McElroy}, {Bottrell}, {Hani}, {Moreno},
  {Croom}, {Hayward}, {Twum}, {Feldmann}, {Hopkins}, {Hernquist}, \&
  {Husemann}}]{2022MNRAS.tmp.1734M}
{McElroy}, R., {Bottrell}, C., {Hani}, M.~H., {et~al.} 2022, \mnras,
  \dodoi{10.1093/mnras/stac1715}

\bibitem[{{McGee} {et~al.}(2011){McGee}, {Balogh}, {Wilman}, {Bower},
  {Mulchaey}, {Parker}, \& {Oemler}}]{2011MNRAS.413..996M}
{McGee}, S.~L., {Balogh}, M.~L., {Wilman}, D.~J., {et~al.} 2011, \mnras, 413,
  996, \dodoi{10.1111/j.1365-2966.2010.18189.x}

\bibitem[{{Naab} \& {Burkert}(2003)}]{2003ApJ...597..893N}
{Naab}, T., \& {Burkert}, A. 2003, \apj, 597, 893, \dodoi{10.1086/378581}

\bibitem[{{Nevin} {et~al.}(2019){Nevin}, {Blecha}, {Comerford}, \&
  {Greene}}]{2019ApJ...872...76N}
{Nevin}, R., {Blecha}, L., {Comerford}, J., \& {Greene}, J. 2019, \apj, 872,
  76, \dodoi{10.3847/1538-4357/aafd34}

\bibitem[{{Nevin} {et~al.}(2021){Nevin}, {Blecha}, {Comerford}, {Greene},
  {Law}, {Stark}, {Westfall}, {Vazquez-Mata}, {Smethurst},
  {Argudo-Fern{\'a}ndez}, {Brownstein}, \& {Drory}}]{2021ApJ...912...45N}
{Nevin}, R., {Blecha}, L., {Comerford}, J., {et~al.} 2021, \apj, 912, 45,
  \dodoi{10.3847/1538-4357/abe2a9}

\bibitem[{{Nikolic} {et~al.}(2004){Nikolic}, {Cullen}, \&
  {Alexander}}]{2004MNRAS.355..874N}
{Nikolic}, B., {Cullen}, H., \& {Alexander}, P. 2004, \mnras, 355, 874,
  \dodoi{10.1111/j.1365-2966.2004.08366.x}

\bibitem[{{Pan} {et~al.}(2019){Pan}, {Lin}, {Hsieh}, {Barrera-Ballesteros},
  {S{\'a}nchez}, {Hsu}, {Keenan}, {Tissera}, {Boquien}, {Dai}, {Knapen},
  {Riffel}, {Argudo-Fern{\'a}ndez}, {Xiao}, \& {Yuan}}]{2019ApJ...881..119P}
{Pan}, H.-A., {Lin}, L., {Hsieh}, B.-C., {et~al.} 2019, \apj, 881, 119,
  \dodoi{10.3847/1538-4357/ab311c}

\bibitem[{{Patton} {et~al.}(2013){Patton}, {Torrey}, {Ellison}, {Mendel}, \&
  {Scudder}}]{2013MNRAS.433L..59P}
{Patton}, D.~R., {Torrey}, P., {Ellison}, S.~L., {Mendel}, J.~T., \& {Scudder},
  J.~M. 2013, \mnras, 433, L59, \dodoi{10.1093/mnrasl/slt058}

\bibitem[{{Pearson} {et~al.}(2019){Pearson}, {Wang}, {Trayford}, {Petrillo}, \&
  {van der Tak}}]{2019A&A...626A..49P}
{Pearson}, W.~J., {Wang}, L., {Trayford}, J.~W., {Petrillo}, C.~E., \& {van der
  Tak}, F.~F.~S. 2019, \aap, 626, A49, \dodoi{10.1051/0004-6361/201935355}

\bibitem[{Pedregosa {et~al.}(2011)Pedregosa, Varoquaux, Gramfort, Michel,
  Thirion, Grisel, Blondel, Prettenhofer, Weiss, Dubourg, Vanderplas, Passos,
  Cournapeau, Brucher, Perrot, \& Duchesnay}]{scikit-learn}
Pedregosa, F., Varoquaux, G., Gramfort, A., {et~al.} 2011, Journal of Machine
  Learning Research, 12, 2825

\bibitem[{{Peng} {et~al.}(2010){Peng}, {Lilly}, {Kova{\v{c}}}, {Bolzonella},
  {Pozzetti}, {Renzini}, {Zamorani}, {Ilbert}, {Knobel}, {Iovino}, {Maier},
  {Cucciati}, {Tasca}, {Carollo}, {Silverman}, {Kampczyk}, {de Ravel},
  {Sanders}, {Scoville}, {Contini}, {Mainieri}, {Scodeggio}, {Kneib}, {Le
  F{\`e}vre}, {Bardelli}, {Bongiorno}, {Caputi}, {Coppa}, {de la Torre},
  {Franzetti}, {Garilli}, {Lamareille}, {Le Borgne}, {Le Brun}, {Mignoli},
  {Perez Montero}, {Pello}, {Ricciardelli}, {Tanaka}, {Tresse}, {Vergani},
  {Welikala}, {Zucca}, {Oesch}, {Abbas}, {Barnes}, {Bordoloi}, {Bottini},
  {Cappi}, {Cassata}, {Cimatti}, {Fumana}, {Hasinger}, {Koekemoer},
  {Leauthaud}, {Maccagni}, {Marinoni}, {McCracken}, {Memeo}, {Meneux}, {Nair},
  {Porciani}, {Presotto}, \& {Scaramella}}]{2010ApJ...721..193P}
{Peng}, Y.-j., {Lilly}, S.~J., {Kova{\v{c}}}, K., {et~al.} 2010, \apj, 721,
  193, \dodoi{10.1088/0004-637X/721/1/193}

\bibitem[{{Rodriguez-Gomez} {et~al.}(2015){Rodriguez-Gomez}, {Genel},
  {Vogelsberger}, {Sijacki}, {Pillepich}, {Sales}, {Torrey}, {Snyder},
  {Nelson}, {Springel}, {Ma}, \& {Hernquist}}]{2015MNRAS.449...49R}
{Rodriguez-Gomez}, V., {Genel}, S., {Vogelsberger}, M., {et~al.} 2015, \mnras,
  449, 49, \dodoi{10.1093/mnras/stv264}

\bibitem[{{Rodriguez-Gomez} {et~al.}(2019){Rodriguez-Gomez}, {Snyder}, {Lotz},
  {Nelson}, {Pillepich}, {Springel}, {Genel}, {Weinberger}, {Tacchella},
  {Pakmor}, {Torrey}, {Marinacci}, {Vogelsberger}, {Hernquist}, \&
  {Thilker}}]{2019MNRAS.483.4140R}
{Rodriguez-Gomez}, V., {Snyder}, G.~F., {Lotz}, J.~M., {et~al.} 2019, \mnras,
  483, 4140, \dodoi{10.1093/mnras/sty3345}

\bibitem[{{S{\'a}nchez} {et~al.}(2012){S{\'a}nchez}, {Kennicutt}, {Gil de Paz},
  {van de Ven}, {V{\'\i}lchez}, {Wisotzki}, {Walcher}, {Mast}, {Aguerri},
  {Albiol-P{\'e}rez}, {Alonso-Herrero}, {Alves}, {Bakos}, {Bart{\'a}kov{\'a}},
  {Bland-Hawthorn}, {Boselli}, {Bomans}, {Castillo-Morales}, {Cortijo-Ferrero},
  {de Lorenzo-C{\'a}ceres}, {Del Olmo}, {Dettmar}, {D{\'\i}az}, {Ellis},
  {Falc{\'o}n-Barroso}, {Flores}, {Gallazzi}, {Garc{\'\i}a-Lorenzo},
  {Gonz{\'a}lez Delgado}, {Gruel}, {Haines}, {Hao}, {Husemann},
  {Igl{\'e}sias-P{\'a}ramo}, {Jahnke}, {Johnson}, {Jungwiert}, {Kalinova},
  {Kehrig}, {Kupko}, {L{\'o}pez-S{\'a}nchez}, {Lyubenova}, {Marino},
  {M{\'a}rmol-Queralt{\'o}}, {M{\'a}rquez}, {Masegosa}, {Meidt},
  {Mendez-Abreu}, {Monreal-Ibero}, {Montijo}, {Mour{\~a}o}, {Palacios-Navarro},
  {Papaderos}, {Pasquali}, {Peletier}, {P{\'e}rez}, {P{\'e}rez}, {Quirrenbach},
  {Rela{\~n}o}, {Rosales-Ortega}, {Roth}, {Ruiz-Lara},
  {S{\'a}nchez-Bl{\'a}zquez}, {Sengupta}, {Singh}, {Stanishev}, {Trager},
  {Vazdekis}, {Viironen}, {Wild}, {Zibetti}, \&
  {Ziegler}}]{2012A&A...538A...8S}
{S{\'a}nchez}, S.~F., {Kennicutt}, R.~C., {Gil de Paz}, A., {et~al.} 2012,
  \aap, 538, A8, \dodoi{10.1051/0004-6361/201117353}

\bibitem[{{Schaye} {et~al.}(2015){Schaye}, {Crain}, {Bower}, {Furlong},
  {Schaller}, {Theuns}, {Dalla Vecchia}, {Frenk}, {McCarthy}, {Helly},
  {Jenkins}, {Rosas-Guevara}, {White}, {Baes}, {Booth}, {Camps}, {Navarro},
  {Qu}, {Rahmati}, {Sawala}, {Thomas}, \& {Trayford}}]{2015MNRAS.446..521S}
{Schaye}, J., {Crain}, R.~A., {Bower}, R.~G., {et~al.} 2015, \mnras, 446, 521,
  \dodoi{10.1093/mnras/stu2058}

\bibitem[{{Scott} \& {Kaviraj}(2014)}]{2014MNRAS.437.2137S}
{Scott}, C., \& {Kaviraj}, S. 2014, \mnras, 437, 2137,
  \dodoi{10.1093/mnras/stt2014}

\bibitem[{{Scudder} {et~al.}(2012){Scudder}, {Ellison}, {Torrey}, {Patton}, \&
  {Mendel}}]{2012MNRAS.426..549S}
{Scudder}, J.~M., {Ellison}, S.~L., {Torrey}, P., {Patton}, D.~R., \& {Mendel},
  J.~T. 2012, \mnras, 426, 549, \dodoi{10.1111/j.1365-2966.2012.21749.x}

\bibitem[{{Smee} {et~al.}(2013){Smee}, {Gunn}, {Uomoto}, {Roe}, {Schlegel},
  {Rockosi}, {Carr}, {Leger}, {Dawson}, {Olmstead}, {Brinkmann}, {Owen},
  {Barkhouser}, {Honscheid}, {Harding}, {Long}, {Lupton}, {Loomis}, {Anderson},
  {Annis}, {Bernardi}, {Bhardwaj}, {Bizyaev}, {Bolton}, {Brewington}, {Briggs},
  {Burles}, {Burns}, {Castander}, {Connolly}, {Davenport}, {Ebelke}, {Epps},
  {Feldman}, {Friedman}, {Frieman}, {Heckman}, {Hull}, {Knapp}, {Lawrence},
  {Loveday}, {Mannery}, {Malanushenko}, {Malanushenko}, {Merrelli}, {Muna},
  {Newman}, {Nichol}, {Oravetz}, {Pan}, {Pope}, {Ricketts}, {Shelden},
  {Sandford}, {Siegmund}, {Simmons}, {Smith}, {Snedden}, {Schneider},
  {SubbaRao}, {Tremonti}, {Waddell}, \& {York}}]{2013AJ....146...32S}
{Smee}, S.~A., {Gunn}, J.~E., {Uomoto}, A., {et~al.} 2013, \aj, 146, 32,
  \dodoi{10.1088/0004-6256/146/2/32}

\bibitem[{{Somerville} \& {Dav{\'e}}(2015)}]{2015ARA&A..53...51S}
{Somerville}, R.~S., \& {Dav{\'e}}, R. 2015, \araa, 53, 51,
  \dodoi{10.1146/annurev-astro-082812-140951}

\bibitem[{{Springel} {et~al.}(2005){Springel}, {Di Matteo}, \&
  {Hernquist}}]{2005MNRAS.361..776S}
{Springel}, V., {Di Matteo}, T., \& {Hernquist}, L. 2005, \mnras, 361, 776,
  \dodoi{10.1111/j.1365-2966.2005.09238.x}

\bibitem[{{Thorp} {et~al.}(2019){Thorp}, {Ellison}, {Simard}, {S{\'a}nchez}, \&
  {Antonio}}]{2019MNRAS.482L..55T}
{Thorp}, M.~D., {Ellison}, S.~L., {Simard}, L., {S{\'a}nchez}, S.~F., \&
  {Antonio}, B. 2019, \mnras, 482, L55, \dodoi{10.1093/mnrasl/sly185}

\bibitem[{{Tomczak} {et~al.}(2016){Tomczak}, {Quadri}, {Tran}, {Labb{\'e}},
  {Straatman}, {Papovich}, {Glazebrook}, {Allen}, {Brammer}, {Cowley},
  {Dickinson}, {Elbaz}, {Inami}, {Kacprzak}, {Morrison}, {Nanayakkara},
  {Persson}, {Rees}, {Salmon}, {Schreiber}, {Spitler}, \&
  {Whitaker}}]{2016ApJ...817..118T}
{Tomczak}, A.~R., {Quadri}, R.~F., {Tran}, K.-V.~H., {et~al.} 2016, \apj, 817,
  118, \dodoi{10.3847/0004-637X/817/2/118}

\bibitem[{{Wake} {et~al.}(2017){Wake}, {Bundy}, {Diamond-Stanic}, {Yan},
  {Blanton}, {Bershady}, {S{\'a}nchez-Gallego}, {Drory}, {Jones}, {Kauffmann},
  {Law}, {Li}, {MacDonald}, {Masters}, {Thomas}, {Tinker}, {Weijmans}, \&
  {Brownstein}}]{2017AJ....154...86W}
{Wake}, D.~A., {Bundy}, K., {Diamond-Stanic}, A.~M., {et~al.} 2017, \aj, 154,
  86, \dodoi{10.3847/1538-3881/aa7ecc}

\bibitem[{{Walmsley} {et~al.}(2019){Walmsley}, {Ferguson}, {Mann}, \&
  {Lintott}}]{2019MNRAS.483.2968W}
{Walmsley}, M., {Ferguson}, A. M.~N., {Mann}, R.~G., \& {Lintott}, C.~J. 2019,
  \mnras, 483, 2968, \dodoi{10.1093/mnras/sty3232}

\bibitem[{{Wetzel} {et~al.}(2013){Wetzel}, {Tinker}, {Conroy}, \& {van den
  Bosch}}]{2013MNRAS.432..336W}
{Wetzel}, A.~R., {Tinker}, J.~L., {Conroy}, C., \& {van den Bosch}, F.~C. 2013,
  \mnras, 432, 336, \dodoi{10.1093/mnras/stt469}

\bibitem[{{Whitaker} {et~al.}(2012){Whitaker}, {van Dokkum}, {Brammer}, \&
  {Franx}}]{2012ApJ...754L..29W}
{Whitaker}, K.~E., {van Dokkum}, P.~G., {Brammer}, G., \& {Franx}, M. 2012,
  \apjl, 754, L29, \dodoi{10.1088/2041-8205/754/2/L29}

\bibitem[{{Whitaker} {et~al.}(2014){Whitaker}, {Franx}, {Leja}, {van Dokkum},
  {Henry}, {Skelton}, {Fumagalli}, {Momcheva}, {Brammer}, {Labb{\'e}},
  {Nelson}, \& {Rigby}}]{2014ApJ...795..104W}
{Whitaker}, K.~E., {Franx}, M., {Leja}, J., {et~al.} 2014, \apj, 795, 104,
  \dodoi{10.1088/0004-637X/795/2/104}

\bibitem[{{White} \& {Frenk}(1991)}]{1991ApJ...379...52W}
{White}, S. D.~M., \& {Frenk}, C.~S. 1991, \apj, 379, 52,
  \dodoi{10.1086/170483}

\bibitem[{{White} \& {Rees}(1978)}]{1978MNRAS.183..341W}
{White}, S.~D.~M., \& {Rees}, M.~J. 1978, \mnras, 183, 341,
  \dodoi{10.1093/mnras/183.3.341}

\bibitem[{{Woods} {et~al.}(2010){Woods}, {Geller}, {Kurtz}, {Westra},
  {Fabricant}, \& {Dell'Antonio}}]{2010AJ....139.1857W}
{Woods}, D.~F., {Geller}, M.~J., {Kurtz}, M.~J., {et~al.} 2010, \aj, 139, 1857,
  \dodoi{10.1088/0004-6256/139/5/1857}

\bibitem[{{Yan} {et~al.}(2016{\natexlab{a}}){Yan}, {Tremonti}, {Bershady},
  {Law}, {Schlegel}, {Bundy}, {Drory}, {MacDonald}, {Bizyaev}, {Blanc},
  {Blanton}, {Cherinka}, {Eigenbrot}, {Gunn}, {Harding}, {Hogg},
  {S{\'a}nchez-Gallego}, {S{\'a}nchez}, {Wake}, {Weijmans}, {Xiao}, \&
  {Zhang}}]{2016AJ....151....8Y}
{Yan}, R., {Tremonti}, C., {Bershady}, M.~A., {et~al.} 2016{\natexlab{a}}, \aj,
  151, 8, \dodoi{10.3847/0004-6256/151/1/8}

\bibitem[{{Yan} {et~al.}(2016{\natexlab{b}}){Yan}, {Bundy}, {Law}, {Bershady},
  {Andrews}, {Cherinka}, {Diamond-Stanic}, {Drory}, {MacDonald},
  {S{\'a}nchez-Gallego}, {Thomas}, {Wake}, {Weijmans}, {Westfall}, {Zhang},
  {Arag{\'o}n-Salamanca}, {Belfiore}, {Bizyaev}, {Blanc}, {Blanton},
  {Brownstein}, {Cappellari}, {D'Souza}, {Emsellem}, {Fu}, {Gaulme}, {Graham},
  {Goddard}, {Gunn}, {Harding}, {Jones}, {Kinemuchi}, {Li}, {Li}, {Maiolino},
  {Mao}, {Maraston}, {Masters}, {Merrifield}, {Oravetz}, {Pan}, {Parejko},
  {Sanchez}, {Schlegel}, {Simmons}, {Thanjavur}, {Tinker}, {Tremonti}, {van den
  Bosch}, \& {Zheng}}]{2016AJ....152..197Y}
{Yan}, R., {Bundy}, K., {Law}, D.~R., {et~al.} 2016{\natexlab{b}}, \aj, 152,
  197, \dodoi{10.3847/0004-6256/152/6/197}

\end{thebibliography}



\end{document}